\documentclass{aastex631}
\usepackage{multirow}

\begin{document}

\title{The electron temperature distribution and the high ionization just behind the shock in the Cygnus Loop}

\author{Masahiro Ichihashi}
\affiliation{Department of Physics, Graduate School of Science, The University of Tokyo,\\
7-3-1 Hongo, Bunkyo-ku, Tokyo 113-0033, Japan}

\author{Aya Bamba}
\affiliation{Department of Physics, Graduate School of Science, The University of Tokyo,\\
7-3-1 Hongo, Bunkyo-ku, Tokyo 113-0033, Japan}
\affiliation{Research Center for the Early Universe, School of Science, The University of Tokyo,\\
7-3-1 Hongo, Bunkyo-ku, Tokyo 113-0033, Japan}
\affiliation{Trans-Scale Quantum Science Institute, The University of Tokyo, \\
7-3-1 Hongo, Bunkyo-ku, Tokyo 113-0033, Japan}

\author{Dai Tateishi}
\altaffiliation{Present address : Collaborative Laboratories for Advanced Decommissioning Science, Japan Atomic Energy Agency, Fukushima, Japan}
\affiliation{Department of Physics, Graduate School of Science, The University of Tokyo,\\
7-3-1 Hongo, Bunkyo-ku, Tokyo 113-0033, Japan}

\author{Kouichi Hagino}
\affiliation{Department of Physics, Graduate School of Science, The University of Tokyo,\\
7-3-1 Hongo, Bunkyo-ku, Tokyo 113-0033, Japan}

\author{Satoru Katsuda}
\affiliation{Graduate School of Science and Engineering, Saitama University,\\
Shimo-Okubo 255, Sakura, Saitama 338-8570, Japan}

\author{Hiroyuki Uchida}
\affiliation{Department of Physics, Kyoto University, \\
Kitashirakawa Oiwake-cho, Sakyo, Kyoto, Kyoto 606-8502, Japan}

\author{Hiromasa Suzuki}
\affiliation{Faculty of Engineering, University of Miyazaki, \\
Miyazaki 889-2192, Japan}

\author{Ryo Yamazaki}
\affiliation{Department of Physical Sciences, Aoyama Gakuin University\\
5-10-1 Fuchinobe Chuo-ku, Sagamihara,
Kanagawa 252-5258, Japan}
\affiliation{Institute of Laser Engineering, Osaka University, \\
2-6 Yamadaoka, Suita, Osaka 565-0871, Japan}

\author{Yutaka Ohira}
\affiliation{Department of Earth and Planetary Science, The University of Tokyo, \\
7-3-1 Hongo, Bunkyo-ku, Tokyo 113-0033, Japan}



\begin{abstract}
The physical processes behind astrophysical collisionless shocks, such as thermal relaxation and ionization after shock passage, remain poorly understood. To investigate these processes, we analyze the northeastern region of the Cygnus Loop with XMM-Newton. The electron temperature is found to increase towards the interior of the remnant ranging from $0.15-0.19\;\mathrm{keV}$ energy range within a spatial scale of $6\;\mathrm{arcmin}$ (or $1.27\;\mathrm{pc}$ 
at a distance of $725\;\mathrm{pc}$) from the shock front. This can be explained well by a modified Sedov solution with radiative cooling. We also show that the ionization timescales determined from our spectroscopy are significantly larger than those estimated based on the electron density of the surrounding materials and the shock velocity. This excess can be qualitatively explained by a mixing of inner multiple plasma components with different ionization states due to turbulence.
\end{abstract}



\section{Introduction} \label{sec:intro}
Collisionless shocks occur in various astrophysical objects with
sufficiently rarefied media, such as supernova remnants \citep{1998ApJS..118..541L}.
In such shocks,
the dissipation of the pre-shock kinetic energy occurs 
on length scales much smaller than the particles'
mean free path for Coulomb collisions. 
As a collisionless shock passes through, plasma is 
compressed and heated. According to the Rankine-Hugoniot relation, 
the expected temperature is $kT=(3/16)mv_s^2$
for sufficiently high Mach number shocks, 
where $m$ and $v_s$ are the particle mass and the shock velocity, respectively. This implies that the electron temperature is lower than that of the proton by a factor of 1836 lower than that of the proton. 
The temperature ratio between electrons and protons changes as the shock velocity slows down due to collisionless shock heating.
Observationally, the electron-to-proton temperature ratio scales as $kT_{\rm e}/kT_{\rm p}\propto v_s^{-2}$ at $2\lesssim M_s\lesssim60$, and finally $kT_{\rm e}/kT_{\rm p}=1$ at $M_s\sim1$ \citep{2015A&A...579A..13V}. 

Downstream of the shock, a relaxation process such as Coulomb collisions \citep{1978ppim.book.....S} slowly equilibrates the proton and electron temperatures. Some clues are observed in Puppis A \citep{2013ApJ...768..182K}, but the relatively large spatial scale of $1\;\mathrm{arcmin}$ (0.4 pc) in this analysis does not allow the detailed comparison with the Coulomb relaxation process. One of the most detailed spatial analysis near the shock front in X-ray studies provided by \cite{2024PASJ...76..800I}. This X-ray study showed that the electron temperature in the northwestern region of SN~1006 is significantly lower than that expected from the Coulomb relaxation. 
However, this result does not necessarily imply that the variation of the electron temperature does not follow the Coulomb relaxation, because some factors affecting the electron temperature distribution, such as the energy leakage to accelerated particles \citep{2003ApJ...589..827B}, are not eliminated.

In this study, we analyze the temperature variation just behind the shock of the northeastern region of the Cygnus Loop. The shock velocity of the Cygnus Loop is $\sim300\;\mathrm{km\;s^{-1}}$ \citep{2009ApJ...702..327S}, much slower than that of  SN~1006, $6000\;\mathrm{km\;s}^{-1}$ \citep{2014ApJ...781...65W}. 
Furthermore, the Cygnus Loop is one of the middle-aged supernova remnants located at the distance $d=725-800\;\mathrm{pc}$ 
from the Earth
\citep[e.g.,][]{2021MNRAS.507..244F,2024MNRAS.528.4490R, 2025ApJ...984...72S}. This small distance from the Earth allows us to perform the detailed spatial analysis near the shock front. 
Some previous studies such as \cite{2007PASJ...59S.163M} and \cite{10.1093/pasj/61.3.503} also analyzed the spatial variation of some plasma parameters with Suzaku data, but they used large spatial divisions ($2.0\;\mathrm{arcmin}$ scale length or $2.3\;\mathrm{arcmin}$ at minimum). Therefore, we use the observational data from XMM-Newton, an X-ray satellite with a large effective area and high spatial resolution, and analyze the spatial distribution of the plasma parameters in more detail than in previous studies.

This paper consists of the following sections: Section~2 describes the observational data and the data reduction we use. Section~3 explains the analysis result of the spatial analysis. Section~4 discusses the spatial distribution of the electron temperature and ionization parameter. Errors represent 90\% confidence intervals.

\section{Observation and Data Reduction} \label{sec:observation}
We use the data from the northeastern edge of the Cygnus Loop observed by X-ray satellite XMM-Newton. The dataset with ObsID 0741820101 is one of the longest exposures (with a resultant exposure time of 101.6 ks) for this region. We use only EPIC MOS2 data to reduce the uncertainties arising from a mixing of the detector background components. The image of the observational data is shown in Figure~\ref{fig:region_all}. These data were reprocessed and analyzed using the XMM-Newton Science Analysis System (SAS) version 21.0.0 and XSPEC software version 12.14.1 \citep{1996ASPC..101...17A}. We use the $\chi^2$ statistic for the analyses described below. 


\section{Analysis and Results} \label{sec:result}

\subsection{Background estimation}
\begin{figure}[ht!]
\includegraphics[width=0.51\linewidth]{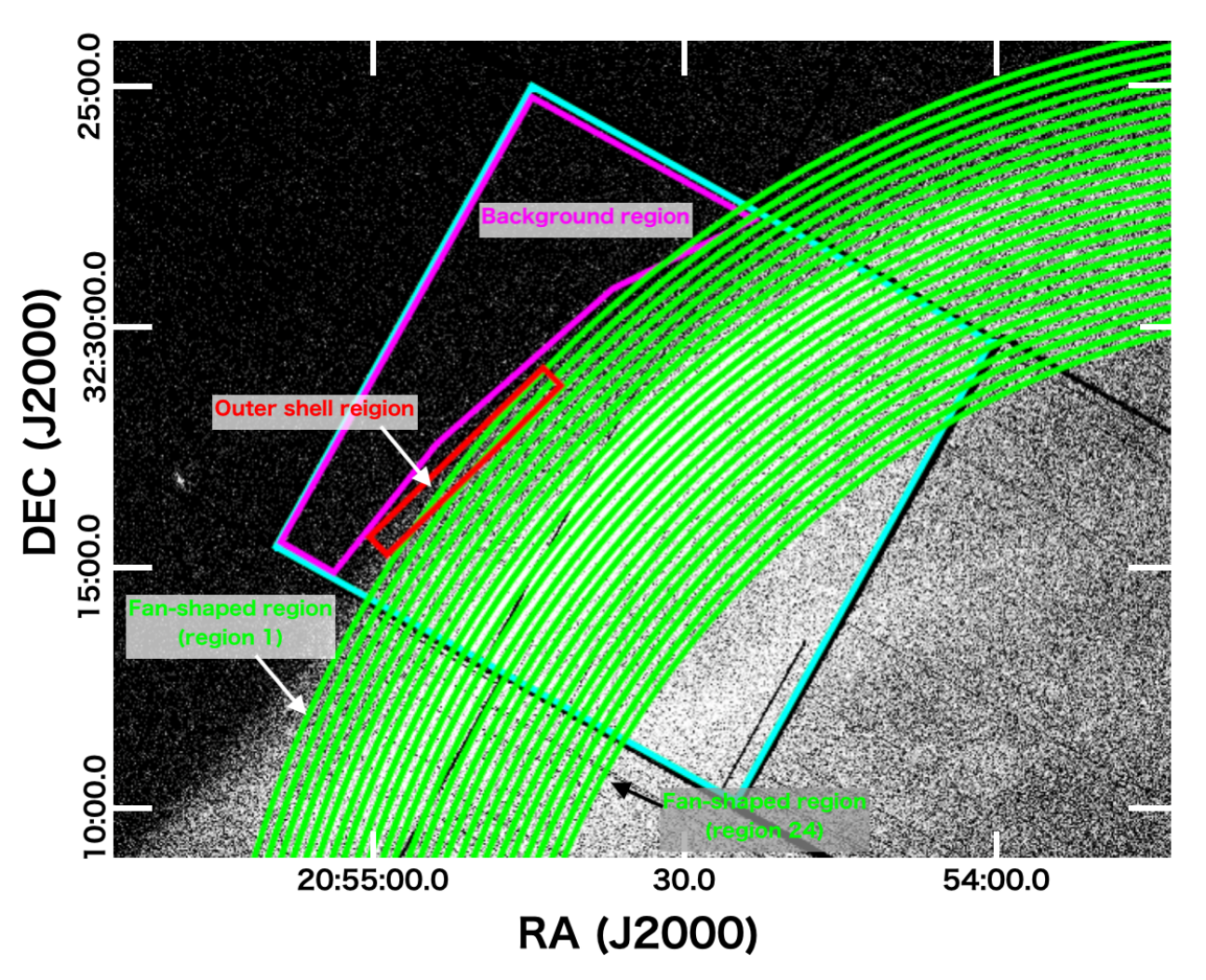}
\includegraphics[width=0.42\linewidth]{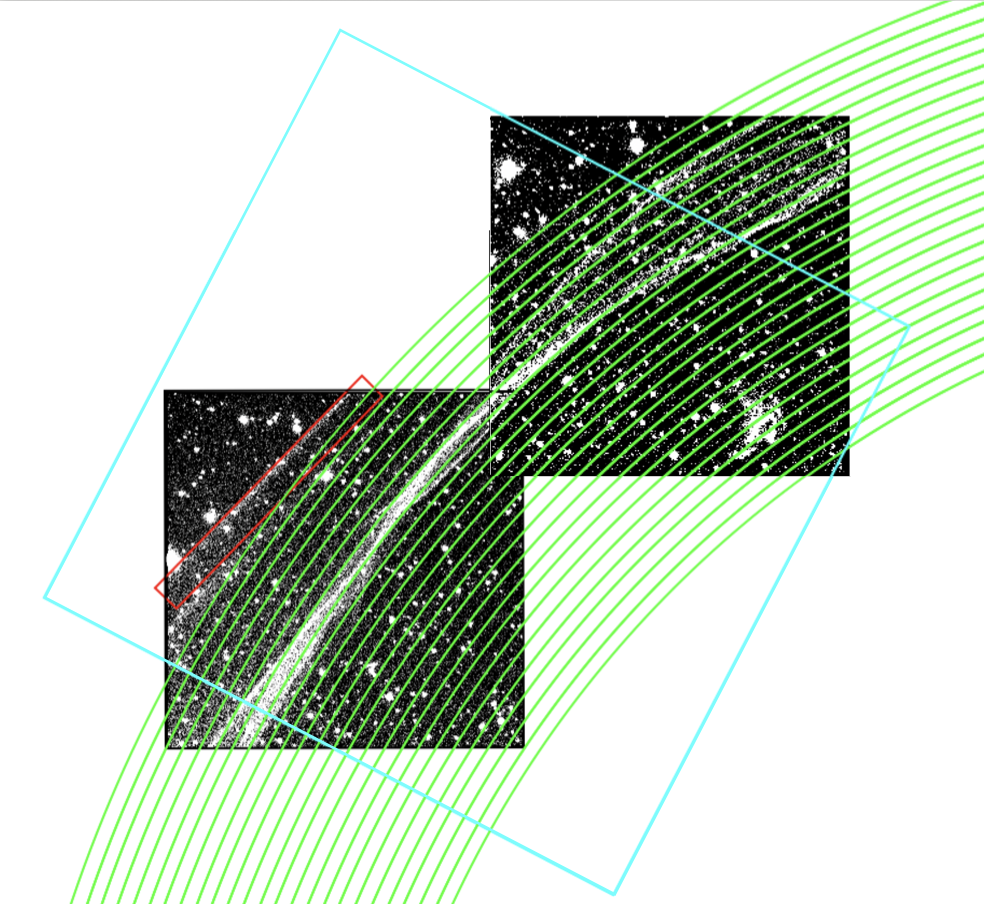}
\caption{The $0.15-12.0\;\mathrm{keV}$ image observed by XMM-Newton EPIC MOS (ObsID:0741820101) (left) 
and the H$_\alpha$ image of the northeastern region of the Cygnus Loop observed at NAO Rozhen by \cite{2023SerAJ.207....9V} (right).
These images are shown on a logarithmic scale. Some shapes in these figures show the regions for spectral analysis below. The cyan square shows the CCD1 of MOS. The magenta polygon shows the background region from which we made the CXB model. Green Fan-shaped regions show the source region. These regions are labeled as layer 1, 2, ..., and 24 from the filament toward downstream. The red box region shows the outer shell region which is located between the two outermost H$_\alpha$ filaments as an indicator of the physical parameters of the shock front.
\label{fig:region_all}}
\end{figure}

The simplest way to estimate the background of the analysis region is to subtract the neighborhood blank-sky spectrum. However, XMM-Newton has a spatial dependence on non-X-ray background \citep{2008A&A...478..575K}, so this method may lead to inaccurate background estimation. For this reason, we consider the background by modeling both the non-X-ray background (NXB) and the cosmic X-ray background (CXB) from a magenta polygon region in Figure~\ref{fig:region_all}.

The NXB model is created using the {\tt mosback} task in the XMM-ESAS package. This task creates a spectrum of NXB of the region we input from the calibration database. We run {\tt mosback} task in for each analyzed region and subtract it from the corresponding source spectrum. Note that the {\tt mosback} task cannot account for some bright instrumental fluorescence lines. In this analysis, Al K$_\alpha$ (1.49 keV) and Si K$_\alpha$ (1.74 keV) are observed as instrumental lines. To account for them, we add two gaussian components. Their central energy, width and normalization are set to free. 

The cosmic X-ray background (CXB) consists of several components. The first is the Cygnus superbubble \citep{2013PASJ...65...14K}, which is modeled as ionization equilibrium collisional plasma ({\tt equil}). The abundance of {\tt equil} is initially fixed at 1 solar, but large residuals remain in the low energy band. These residuals are reduced by freeing the abundances of N, O, Ne and Fe independently.  
The second is active galactic nuclei, which is modeled as a {\tt powerlaw} model \citep{2003ApJ...588..696M}. The power-law index is set to free. As for the interstellar absorption, we use a {\tt tbabs} model with free hydrogen column density. All free parameters are determined by fitting the spectrum of the background region. 

\begin{figure}[ht!]
\plotone{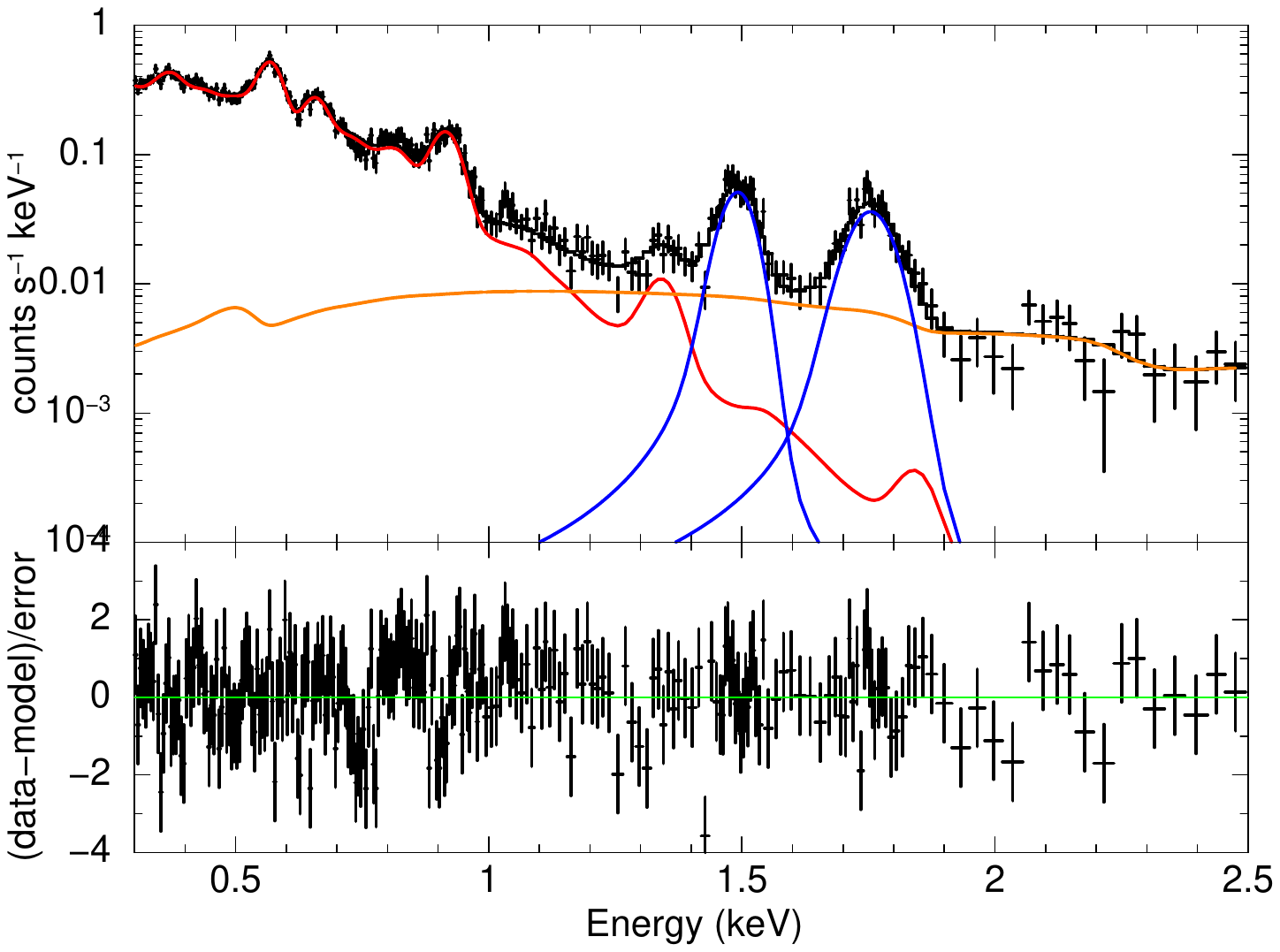}
\caption{The spectrum and the best-fit model of the background region. NXB spectrum made by {\tt mosback} is already subtracted. The red, orange and blue line show the model of the Cygnus supper bubble, active galactic nuclei and NXB fluorescent lines, respectively.\label{fig:best-fit_bkg}}
\end{figure}

The best-fit parameters and the models of CXB are shown in Table~\ref{tab:bkg_parmeter} and Figure~\ref{fig:best-fit_bkg}. 
The best-fit value of the power-law index falls within the range given in the value in \cite{2013PASJ...65...14K}. The best-fit hydrogen column density is an intermediate value between that of previous studies: $2.2\times10^{21}\;\mathrm{cm^{-2}}$ from ROSAT observation of the Cygnus superbubble \citep{2013PASJ...65...14K}, and about $3\times10^{19}\;\mathrm{cm^{-2}}$ from the Dwingeloo H$_\mathrm{I}$ survey \citep{2001A&A...371..675U}.  
When applying the background model to the source spectrum, we fix all the CXB parameters except for the normalization to the values in Table~\ref{tab:bkg_parmeter}. Only the normalization of the background model is kept as a free parameter. The normalization ratio of CXB is fixed to that of the best-fit parameter in Table~\ref{tab:bkg_parmeter}. The width of the Al K$_\alpha$ line is too small and cannot be distinguished from 0, so we set its width to 0 when applying it as an NXB model to the source spectrum.

\begin{table}[ht]
    \centering
    \caption{Best-fit parameter for NXB}
    \label{tab:bkg_parmeter}
    \begin{tabular}{ccc}
        \hline
         Model&Parameter&Value  \\
         \hline
         tbabs & $N_\mathrm{H}$($\times10^{20}\;\mathrm{cm^{-2}}$)&$5.9_{-0.6}^{+0.6}$\\
         vequil&kT(keV)&$0.16_{-0.01}^{+0.01}$\\
         &N&$0.24_{-0.07}^{+0.08}$\\
         &O&$0.24_{-0.02}^{+0.02}$\\
         &Ne&$0.64_{-0.08}^{+0.09}$\\
         &Fe&$1.3_{-0.4}^{+0.6}$\\
         &norm ($\mathrm{cm}^{-2}$)&$4.8_{-0.6}^{+0.7}\times10^{-3}$\\
         \multirow{2}{*}{powerlaw}&index&$1.5_{-0.4}^{+0.4}$\\
         &norm (photon/keV/cm$^2$/s)&$3.3_{-0.8}^{+0.8}\times10^{-5}$\\
         gaussian (Al K$_\alpha$)&E (keV)&$1.5_{-0.1}^{+0.1}$\\
         &sigma (keV)&$<0.01$\\
         &norm (photon/keV/cm$^2$/s)\footnote{at 1 keV\label{fot:at1}}&$4.5_{-0.5}^{+0.5}$\\
         gaussian (Si K$_\alpha$)&E (keV)&$1.8_{-0.1}^{+0.1}$\\
         &sigma (keV)&$2.6_{-1.1}^{+0.9}\times10^{-2}$\\
         &norm (photon/keV/cm$^2$/s)\footref{fot:at1} &$4.2_{-0.5}^{+0.5}\times10^{-3}$\\
         \hline
         \multicolumn{2}{c}{$\chi^2$/d.o.f}&402.65/272\\
         \hline
    \end{tabular}
\end{table}

\subsection{Spectral analysis on the source regions}
To analyze the plasma inside the shock, it is necessary to identify  the location of the shock front. In many cases, the location of an $\mathrm{H}_\alpha$ filament is considered to be the shock front. However, some $\mathrm{H}_\alpha$ filaments overlap in the northeastern region of the Cygnus Loop (\cite{2023SerAJ.207....9V} and the reference therein, see the right figure of Figure \ref{fig:region_all}). This overlap is thought to be due to the projection of the wavy shock front, so the location of the shock front is ambiguous in our analysis region. 
On the other hand, the region inside the brightest $\mathrm{H}_\alpha$ filament, which is located in the same position of the inner discontinuity in the XMM-Newton image, contains no $\mathrm{H}_\alpha$ filaments. This means that this region is certainly inside of all the shock fronts and thus is a reasonable region for analyzing  the plasma inside the shock. Since this filament is arc-shaped, we divide the northeastern region of the Cygnus Loop into fan-shaped sectors, settingone boundary along the filament to analyze the radial profile of plasma parameters. 
Each region has a thickness of 15 arcsec (or $0.05 \;\mathrm{pc}$). We place 8 regions outside the filament and 16 regions inside. In addition, we extract a box region of 300 arcsec width and 30 arcsec thickness, which we call the outer shell region. The boundary of this box region is along two outermost $\mathrm{H}_\alpha$ filaments, so the emission in this region is really from just behind the shock. The spectrum from this region serves as a good indicator of the physical parameters at the shock front. 

We assume the emission is mainly from the shocked interstellar medium heated by the shock wave, so we use an absorbed non-equilibrium ionization collisional plasma model ({\tt nei}). We use the abundance derived from \cite{1989GeCoA..53..197A}. The abundances of C, N, O, Ne, Mg, and Fe are set to be free. Other abundances are fixed to the solar value. The XSPEC model does not include some Fe L lines \citep{2019PASJ...71...61S}, so we add a Gaussian component with center energy of $1.23\;\mathrm{keV}$ and fixed width of 0 to account for them. The interstellar absorption is modeled with the {\tt tbabs} model with the hydrogen column density fixed at $6.0 \times 10^{20}\;\mathrm{cm^{-2}}$. This value is based on \cite{2007PASJ...59S.163M}, which analyzed the same region in Suzaku data and shows that the hydrogen column density is constant in our analysis region. 

\begin{figure}[htp]
\gridline{\fig{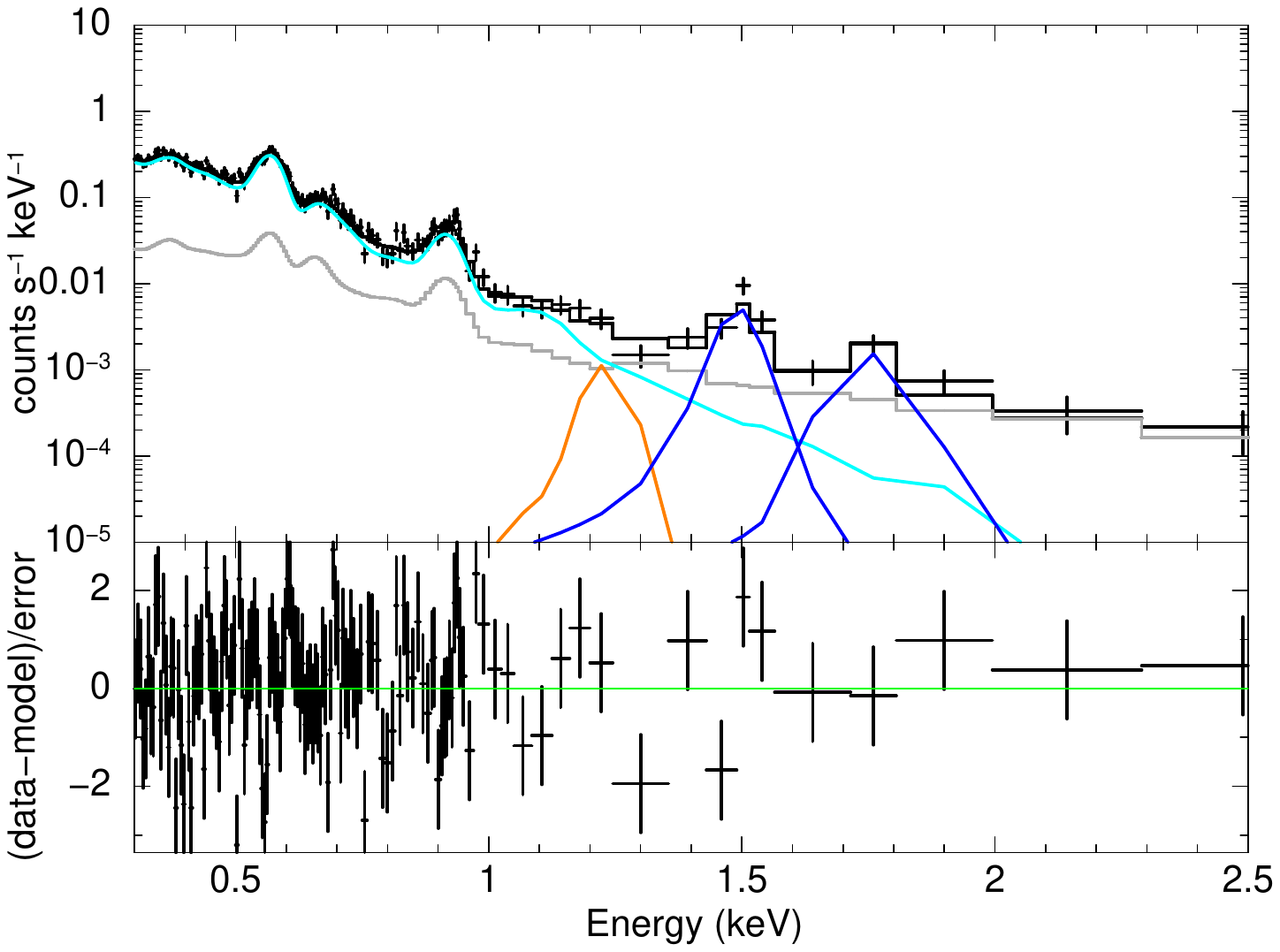}{0.18\textwidth}{layer 1}
          \fig{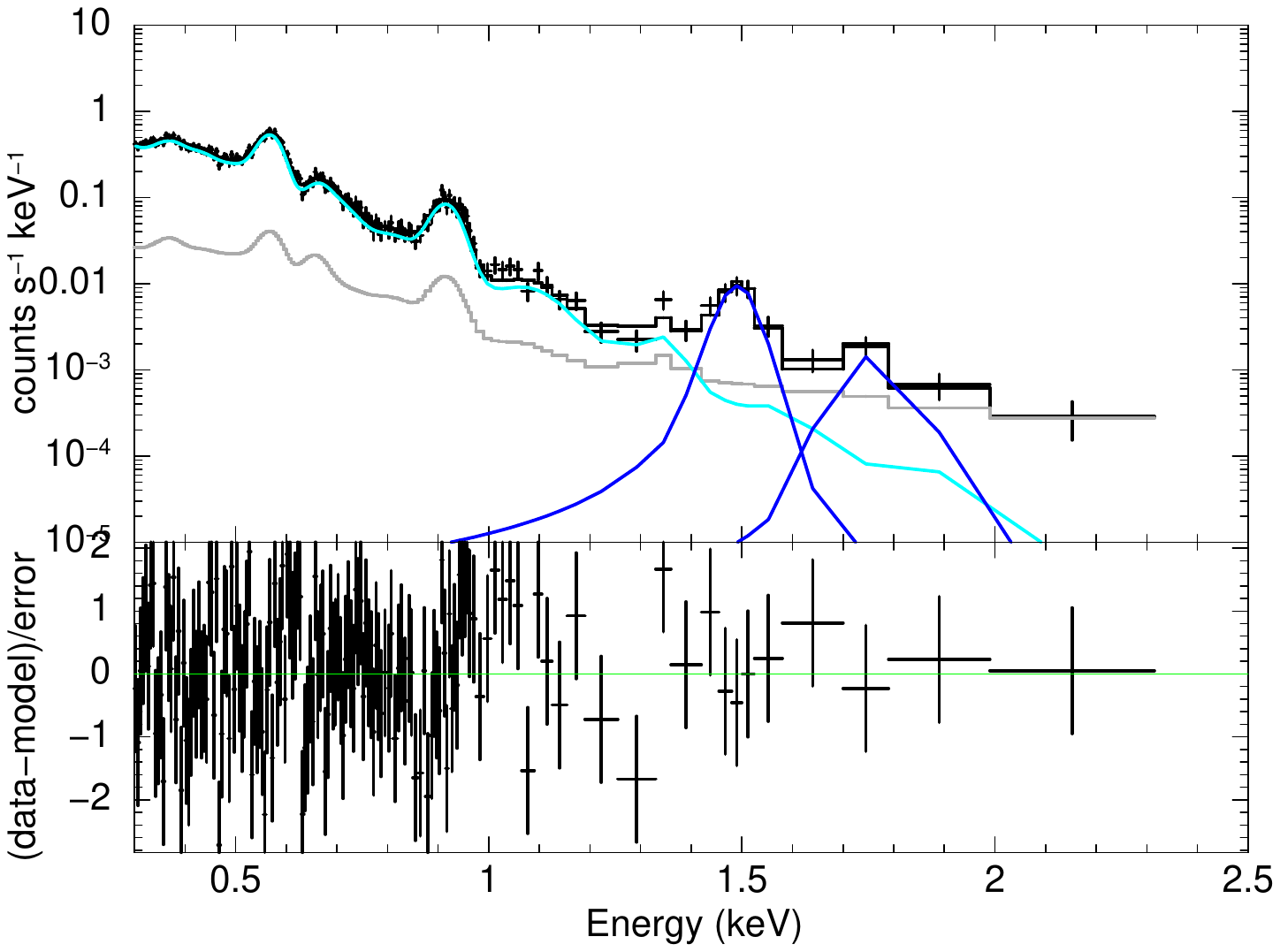}{0.18\textwidth}{layer 2}
          \fig{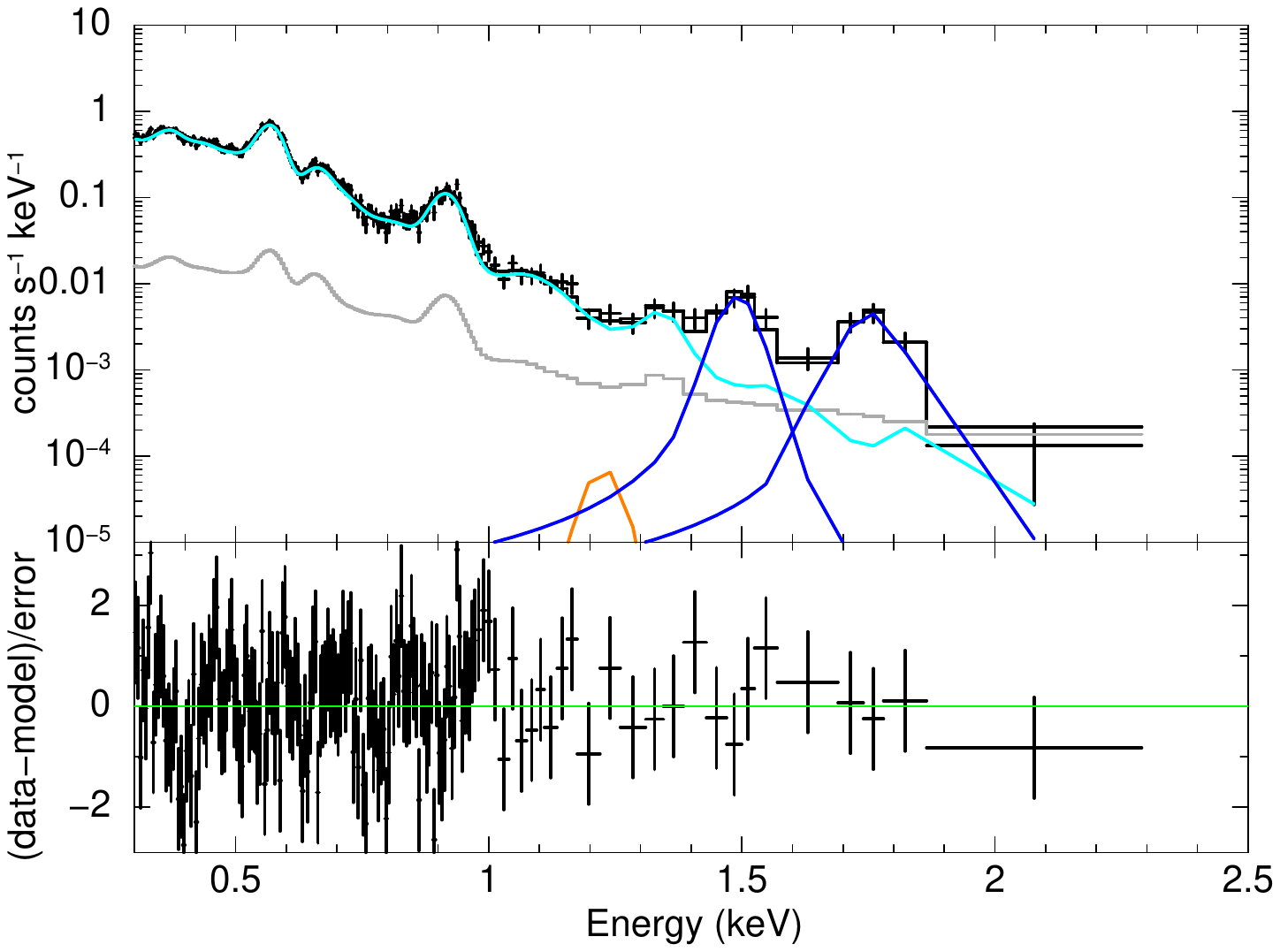}{0.18\textwidth}{layer 3}
          \fig{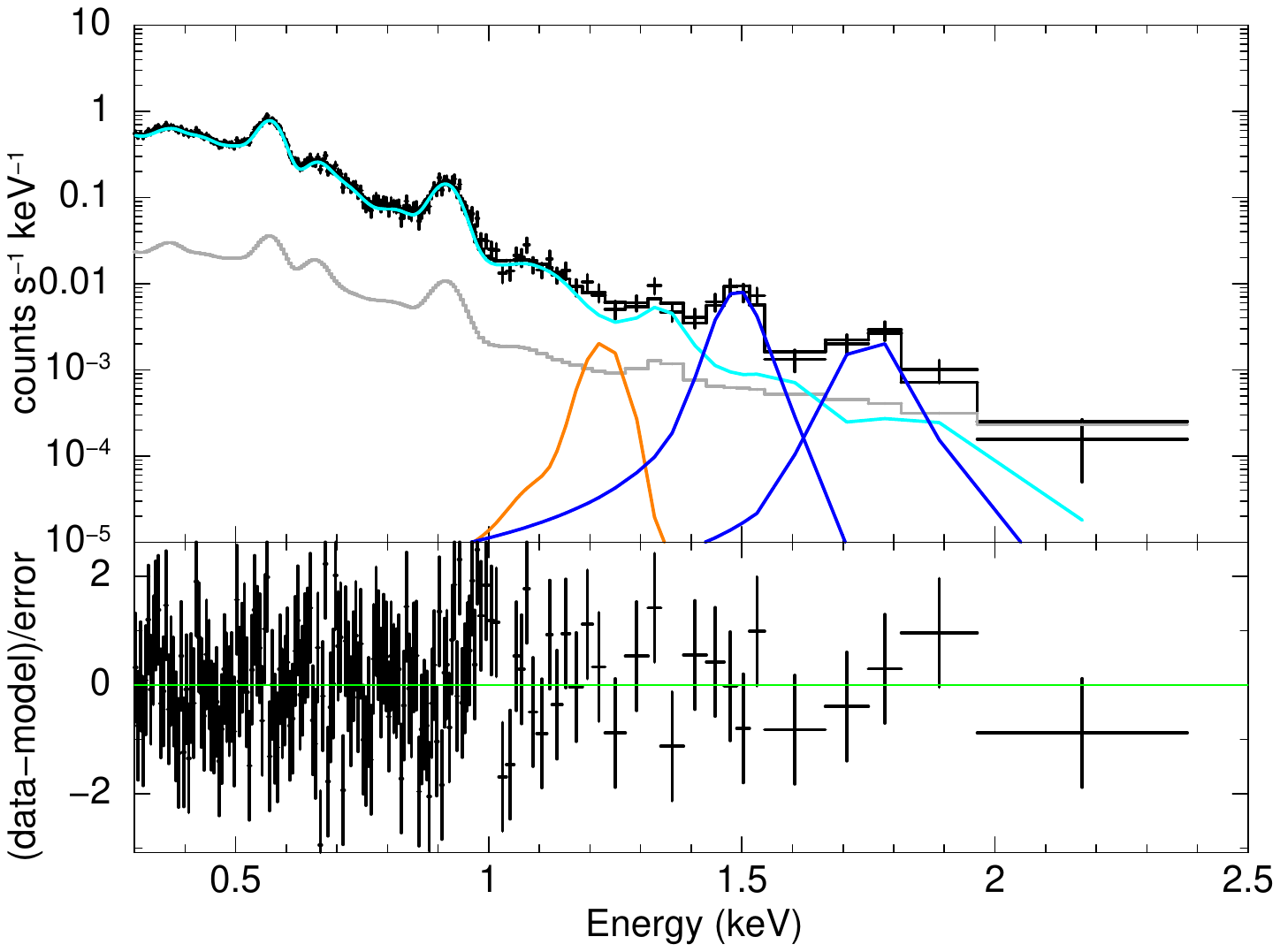}{0.18\textwidth}{layer 4}}
\gridline{\fig{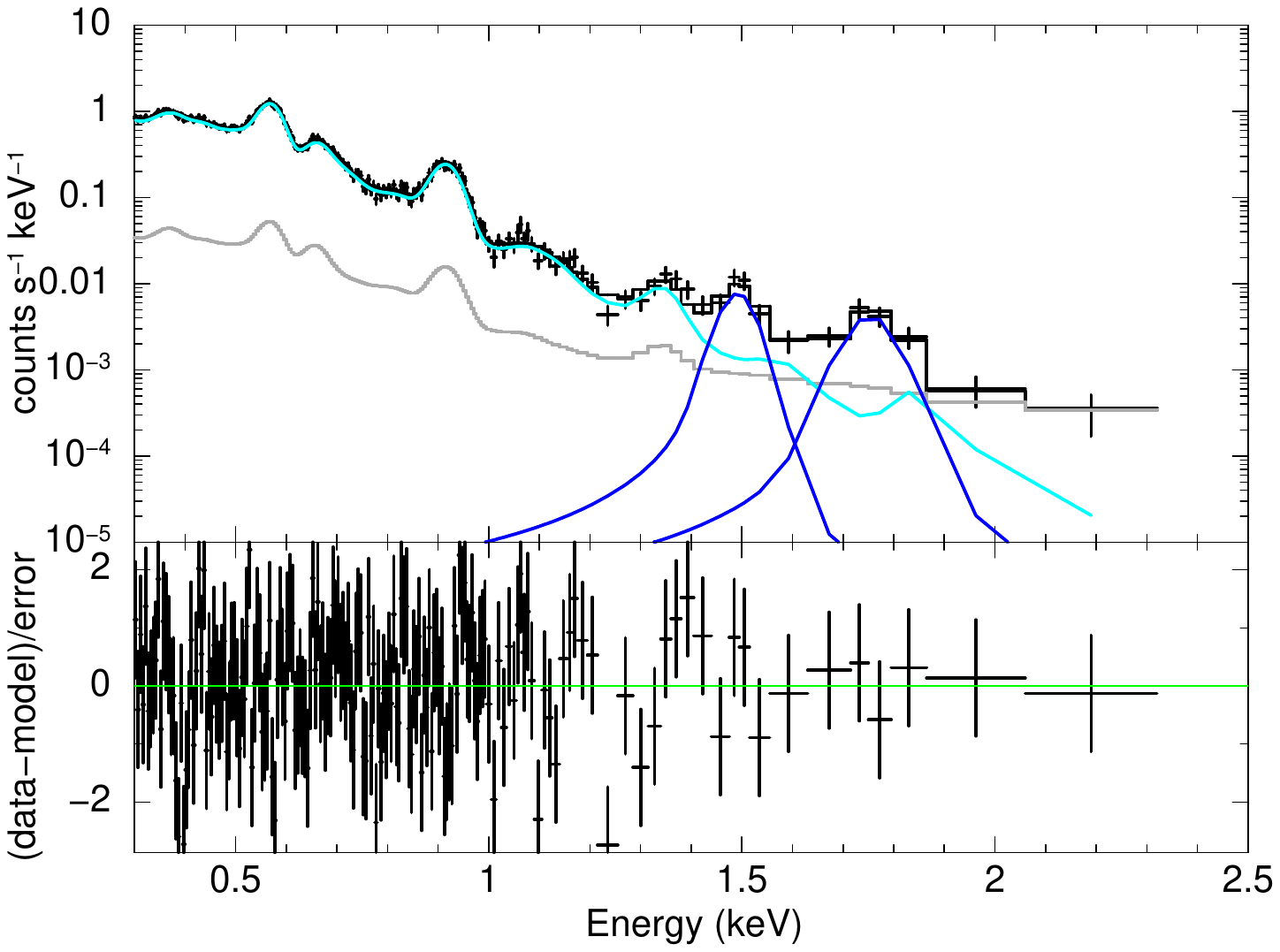}{0.18\textwidth}{layer 5}
          \fig{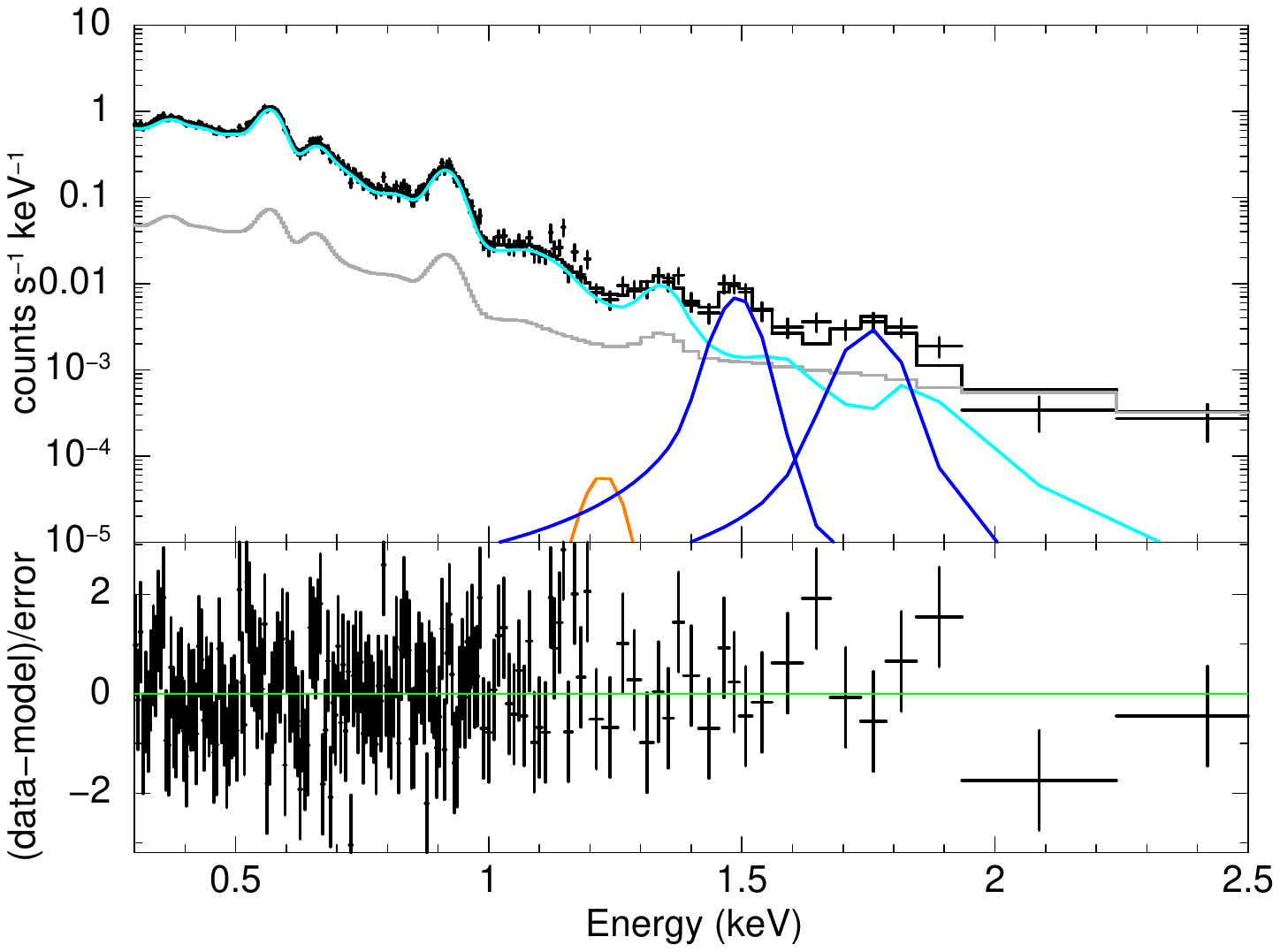}{0.18\textwidth}{layer 6}
          \fig{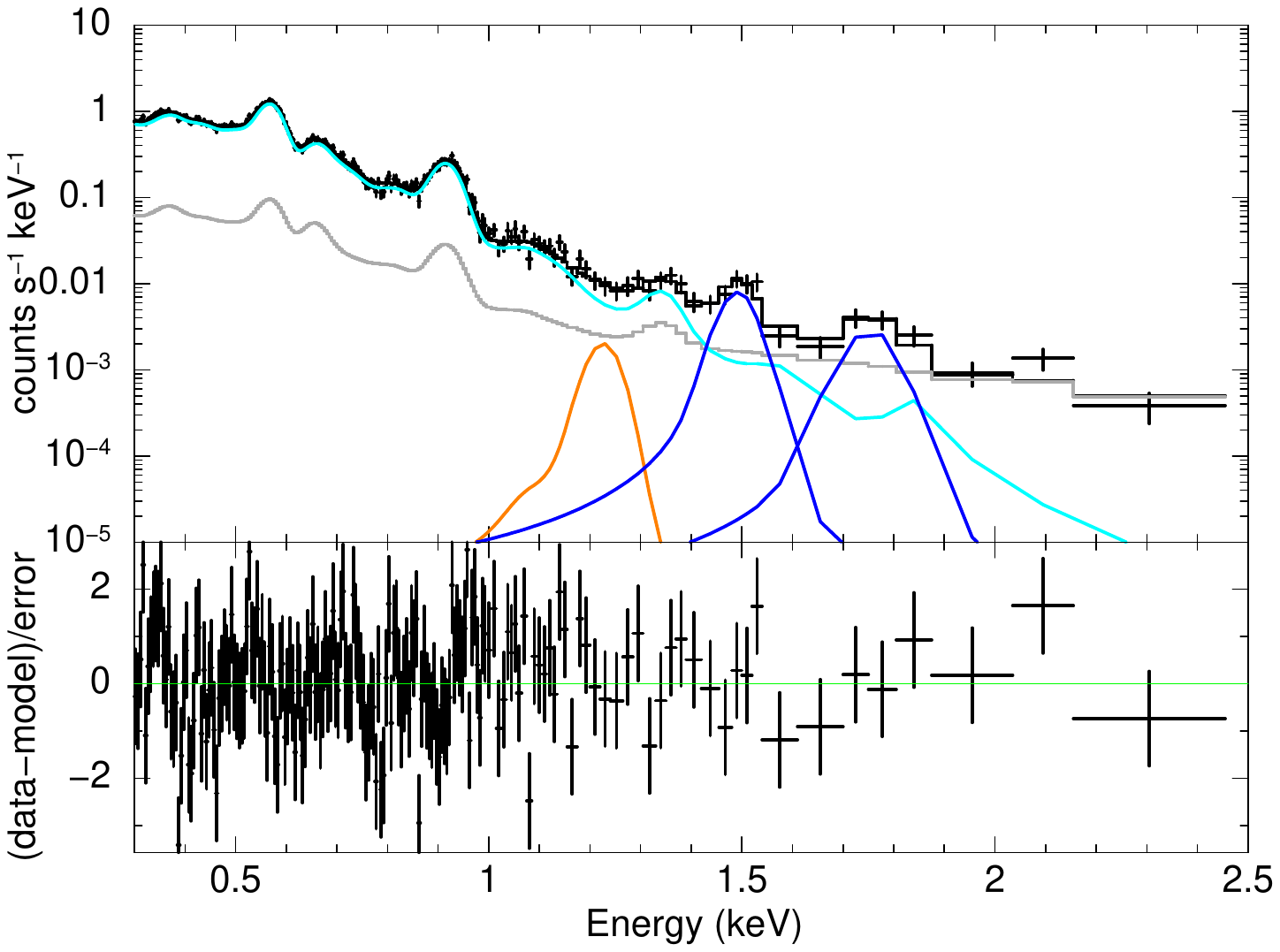}{0.18\textwidth}{layer 7}
          \fig{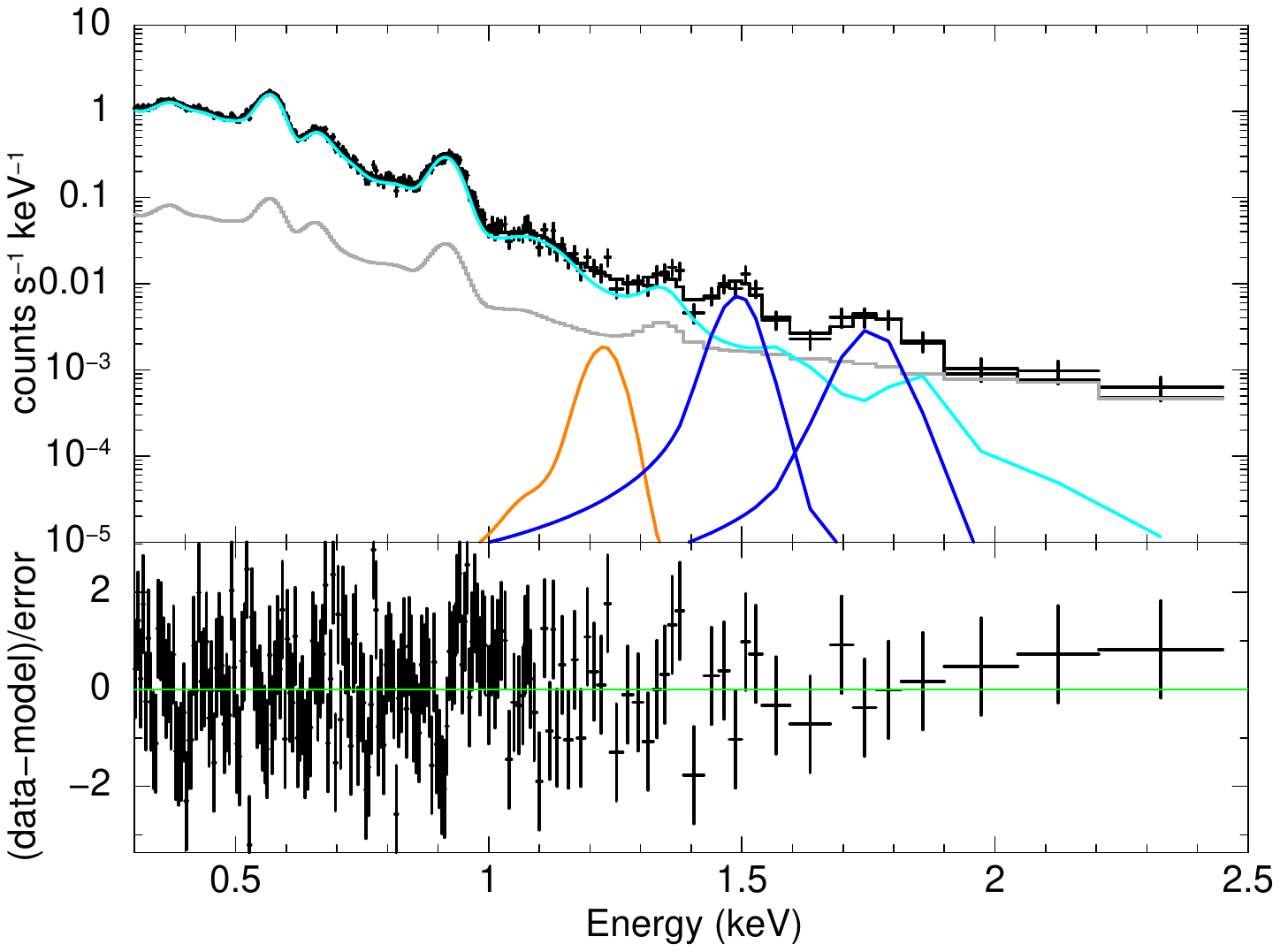}{0.18\textwidth}{layer 8}}
\gridline{\fig{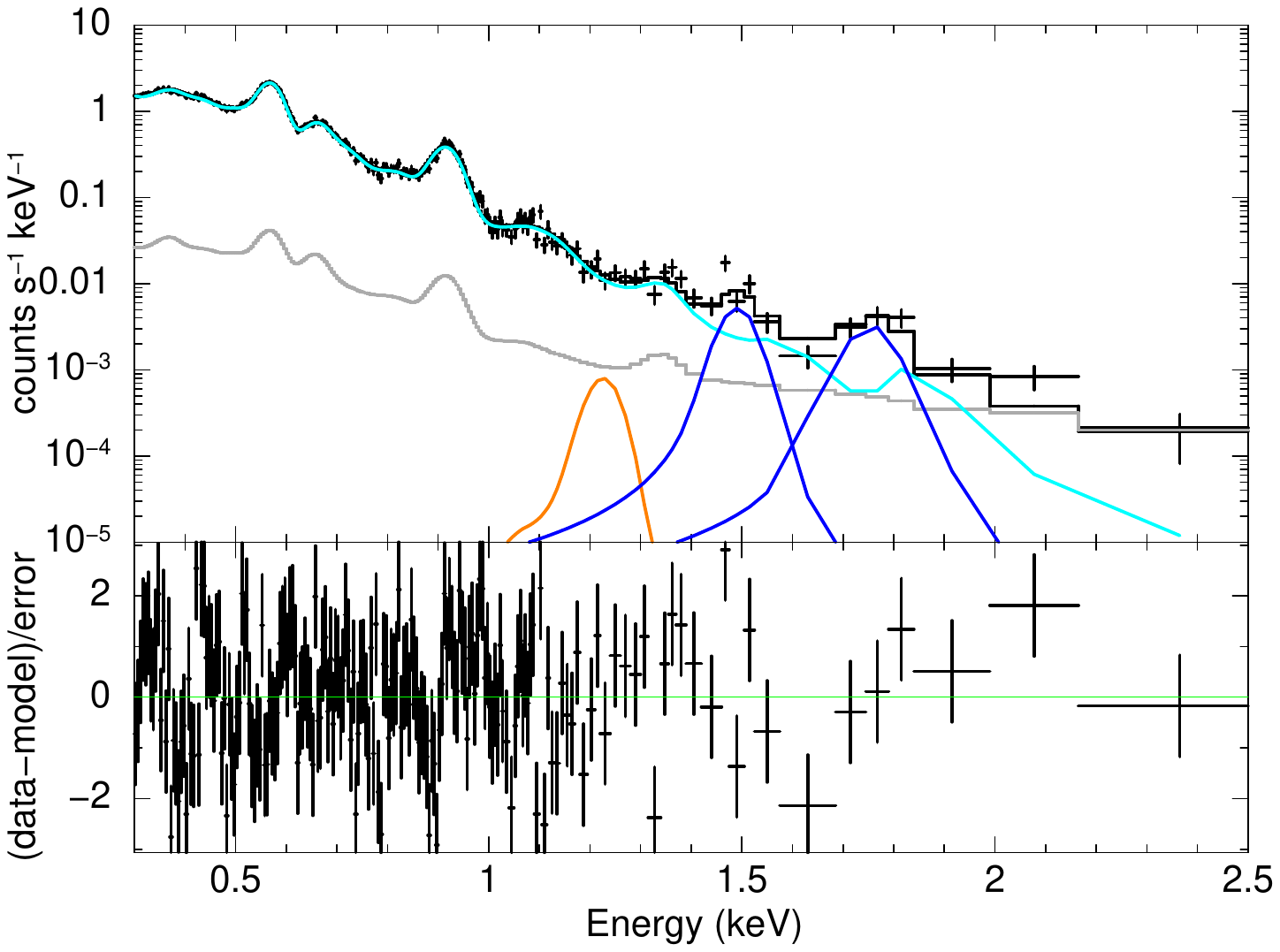}{0.18\textwidth}{layer 9}
          \fig{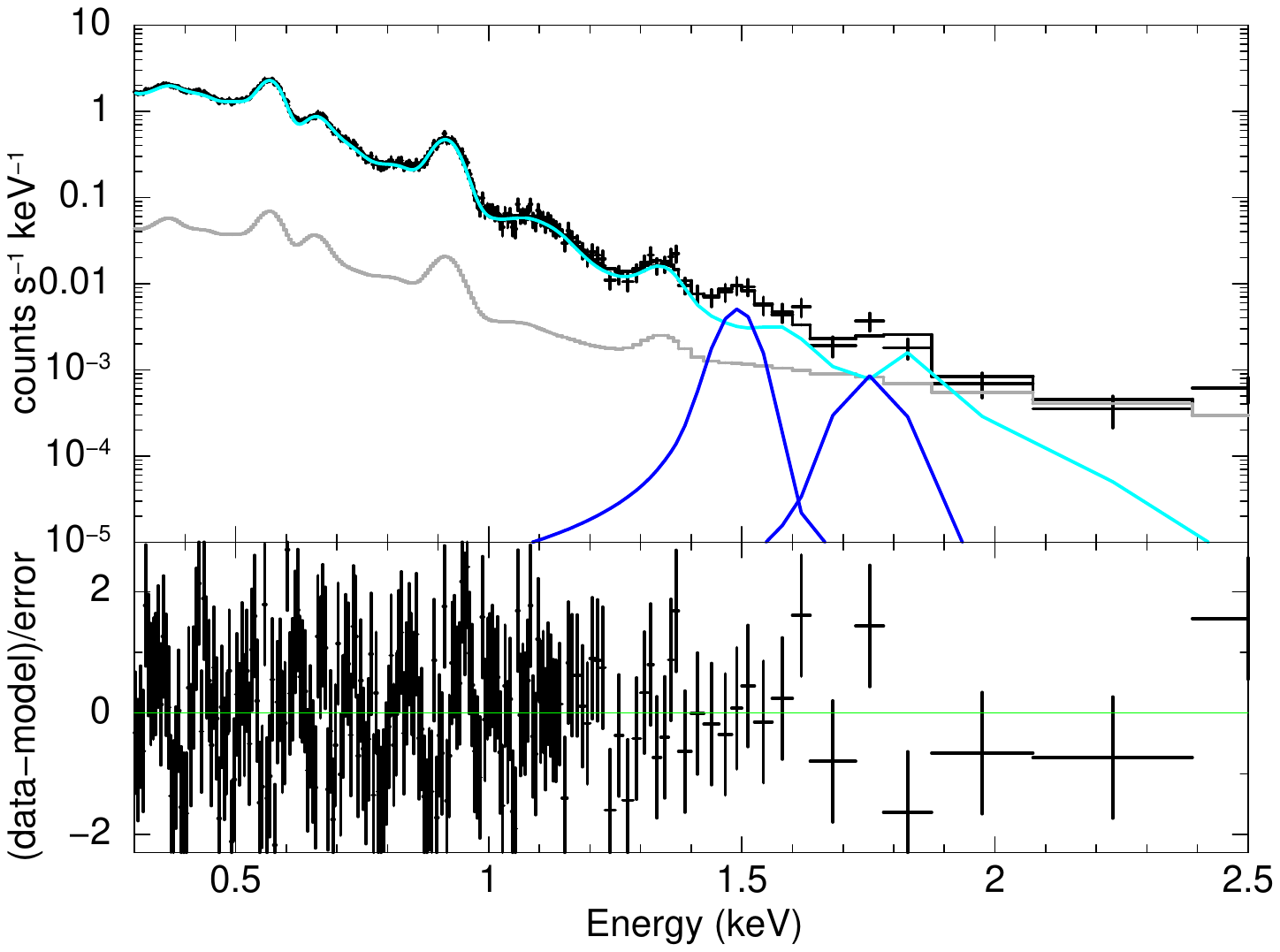}{0.18\textwidth}{layer 10}
          \fig{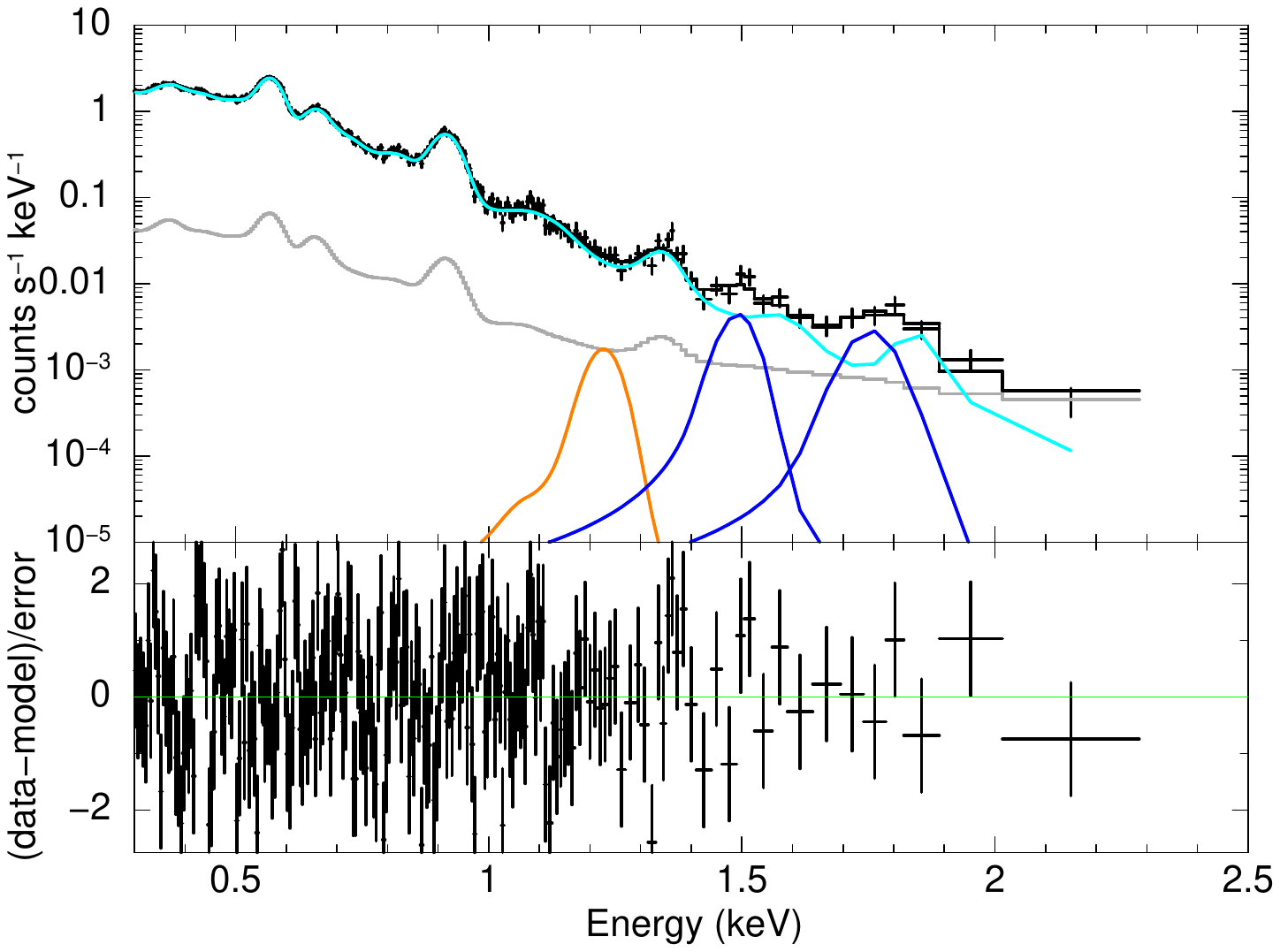}{0.18\textwidth}{layer 11}
          \fig{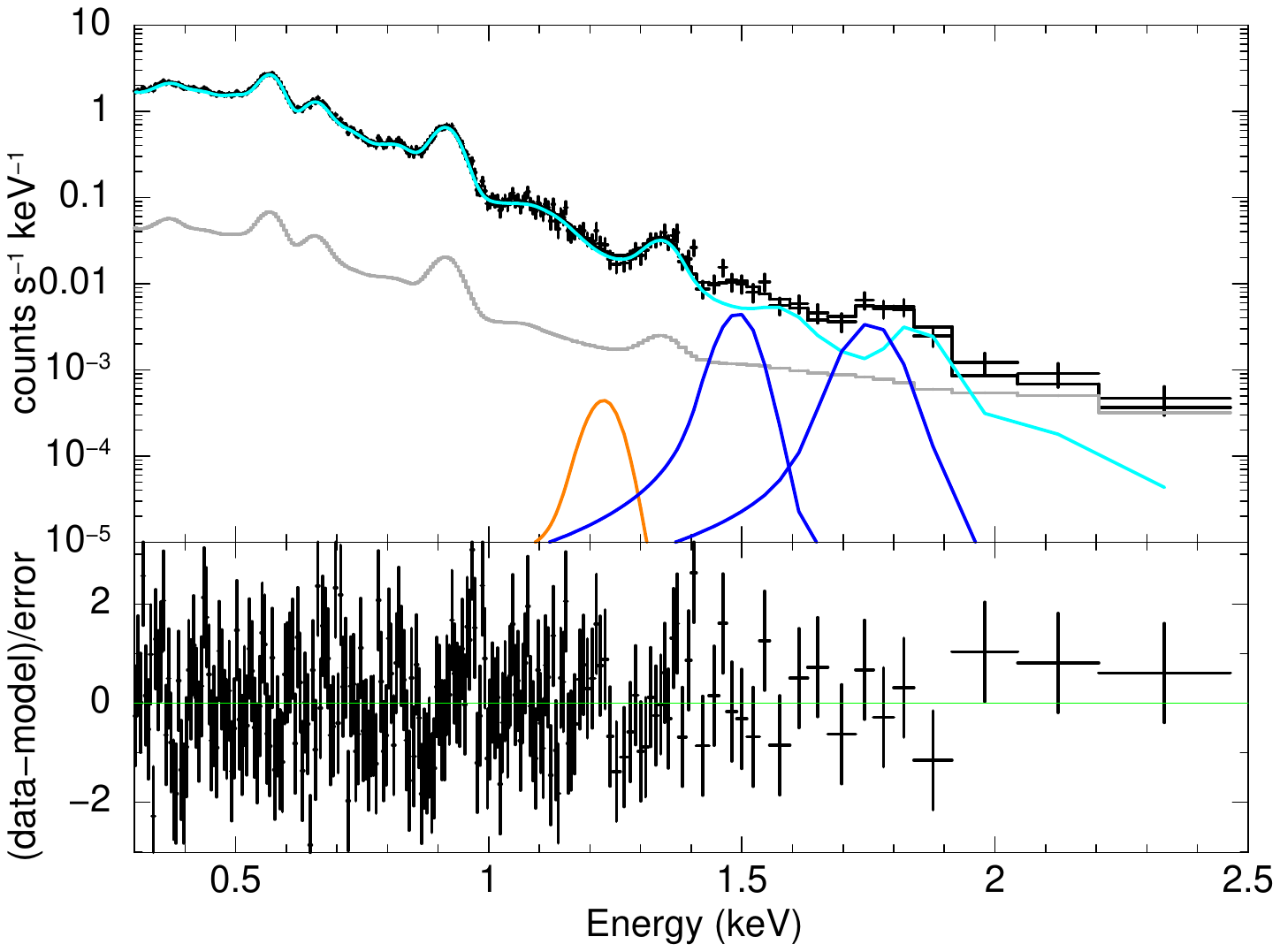}{0.18\textwidth}{layer 12}}
\gridline{\fig{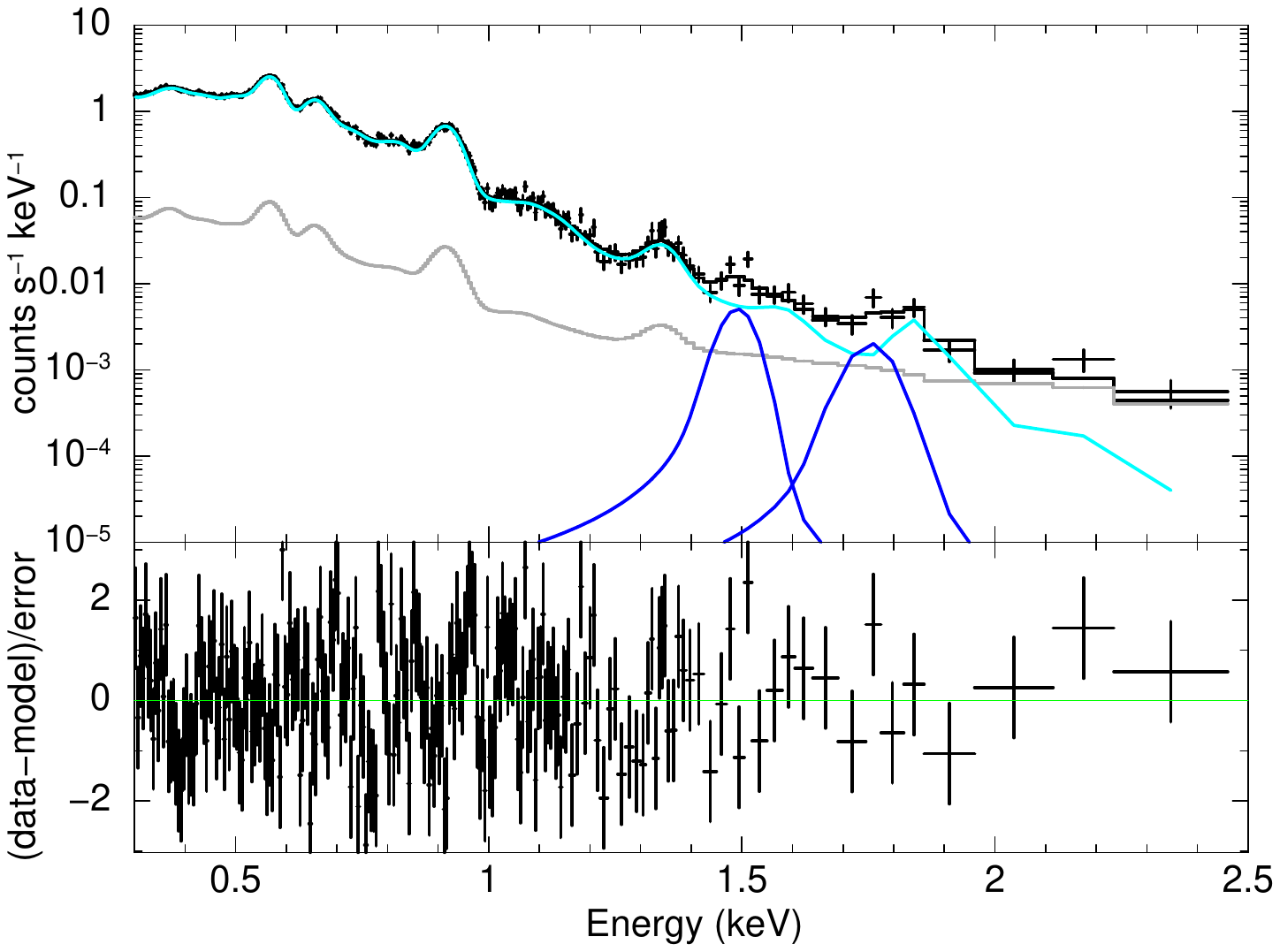}{0.18\textwidth}{layer 13}
          \fig{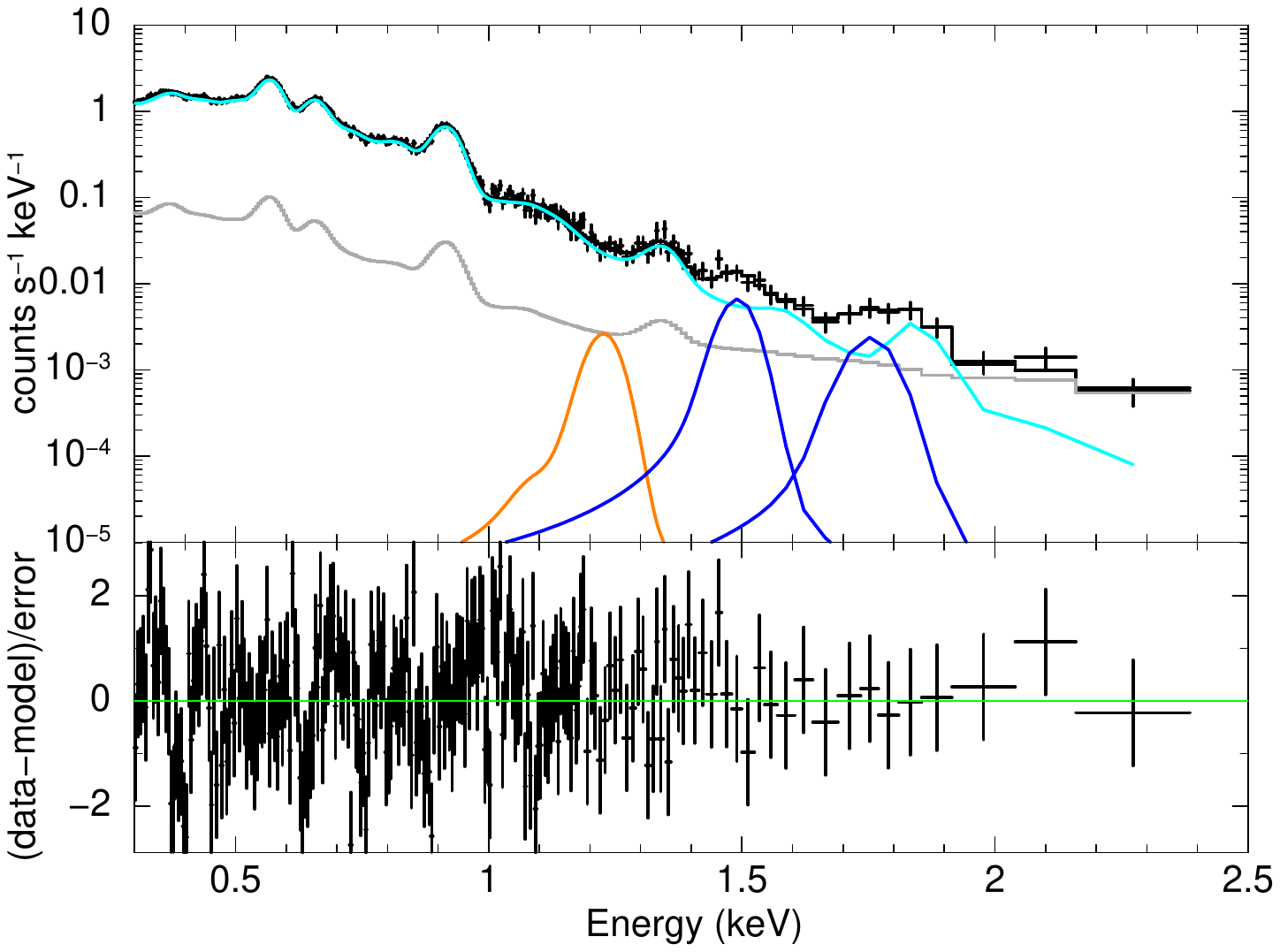}{0.18\textwidth}{layer 14}
          \fig{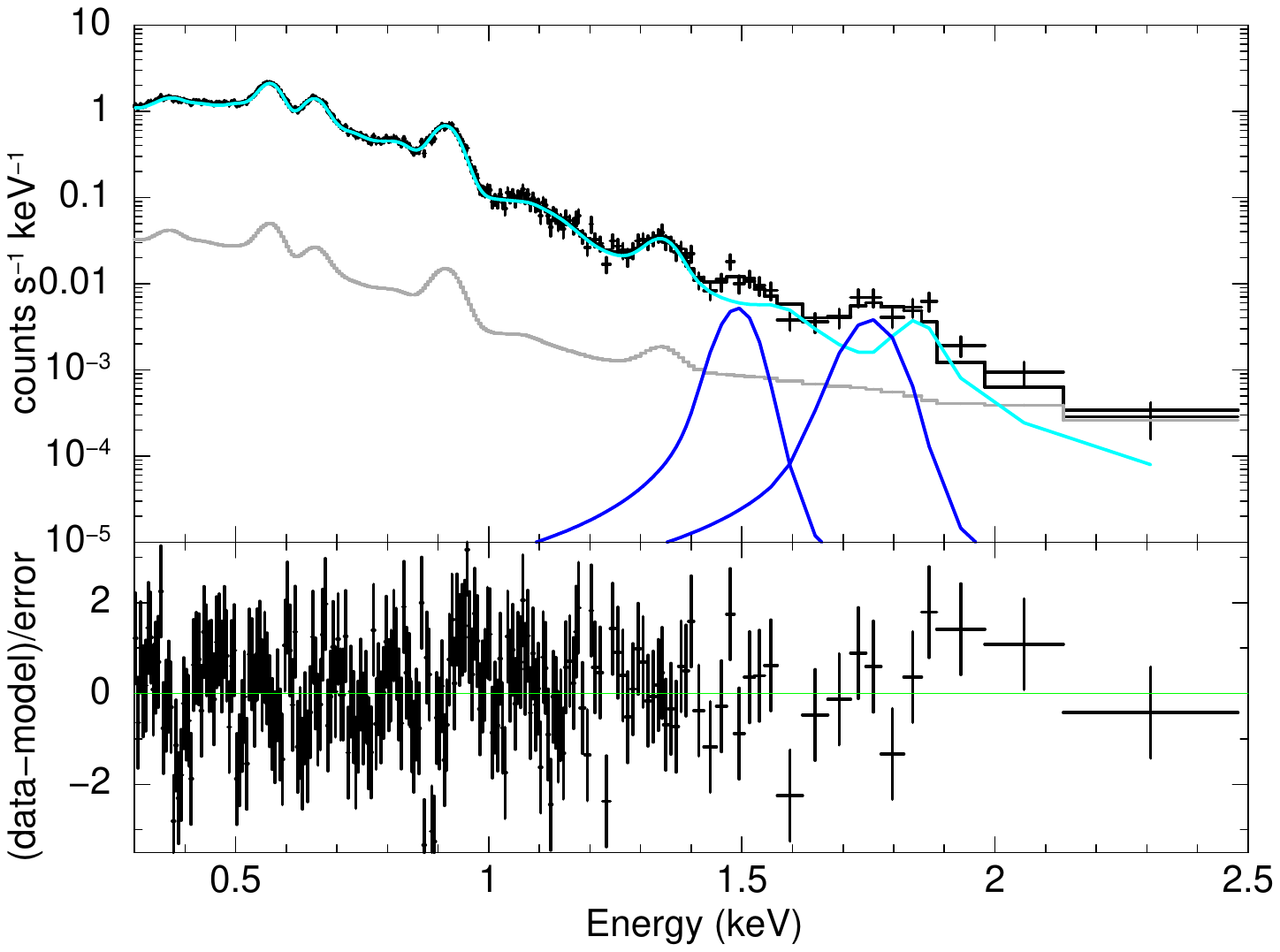}{0.18\textwidth}{layer 15}
          \fig{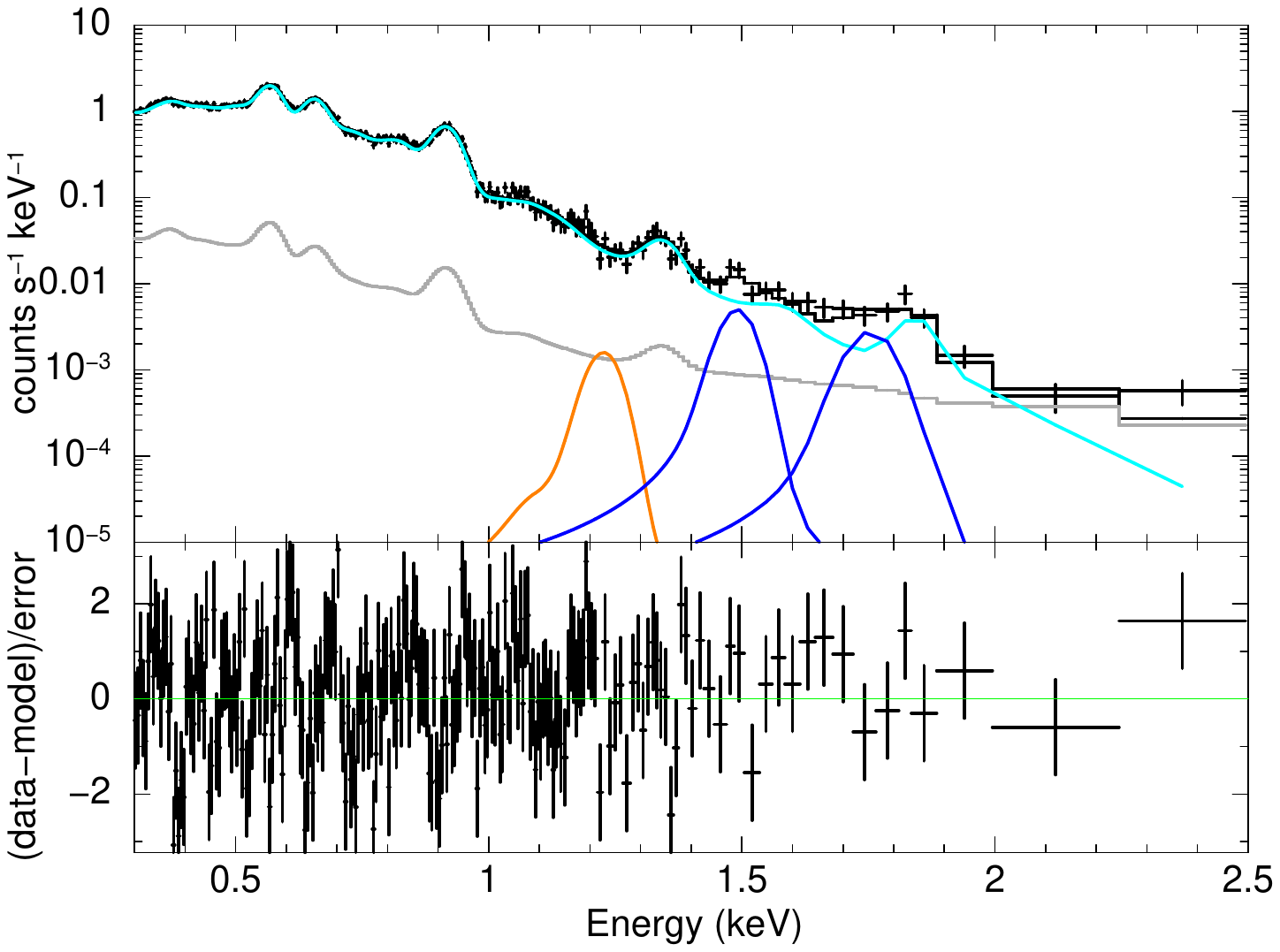}{0.18\textwidth}{layer 16}}
\gridline{\fig{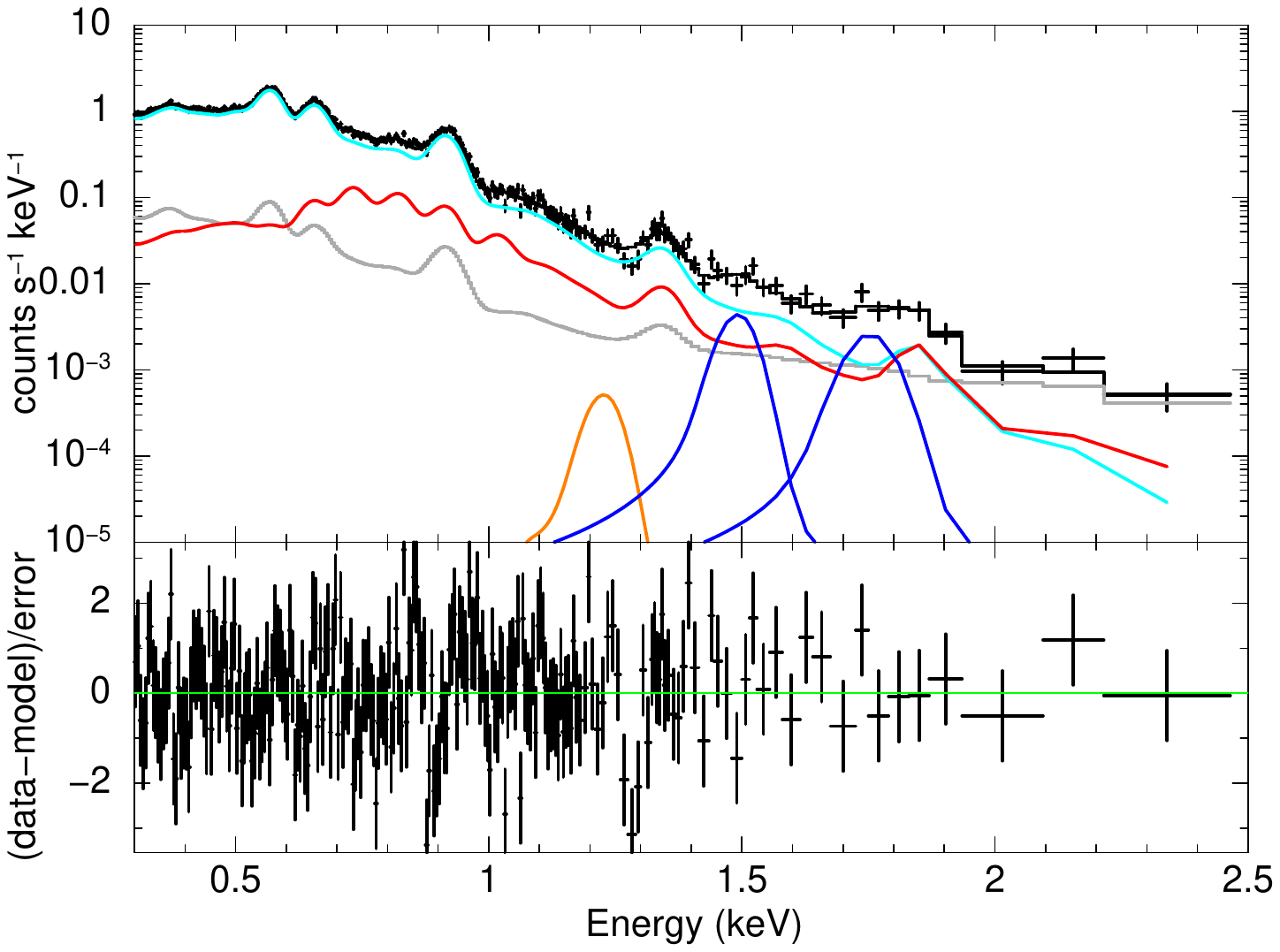}{0.18\textwidth}{layer 17}
          \fig{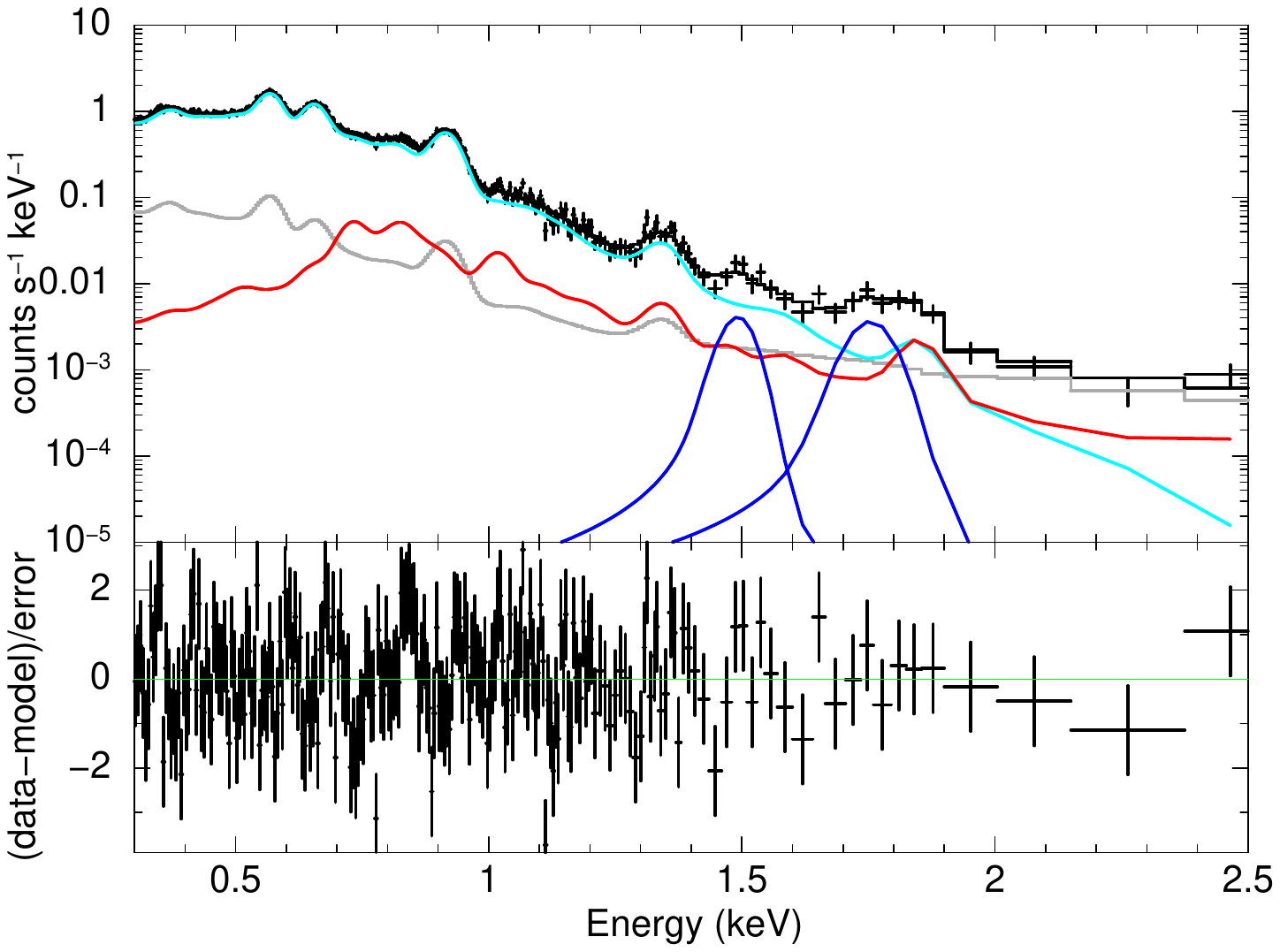}{0.18\textwidth}{layer 18}
          \fig{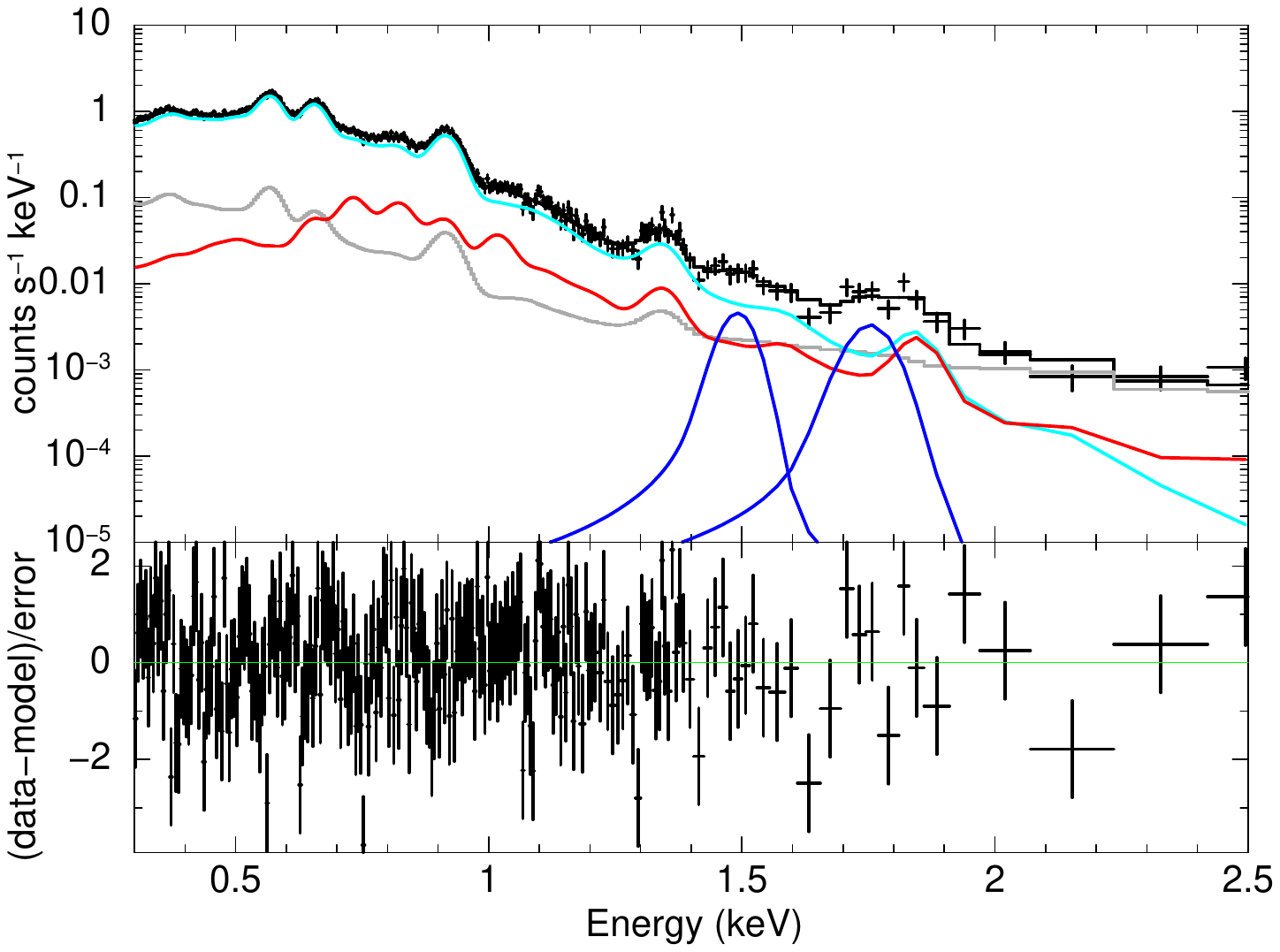}{0.18\textwidth}{layer 19}
          \fig{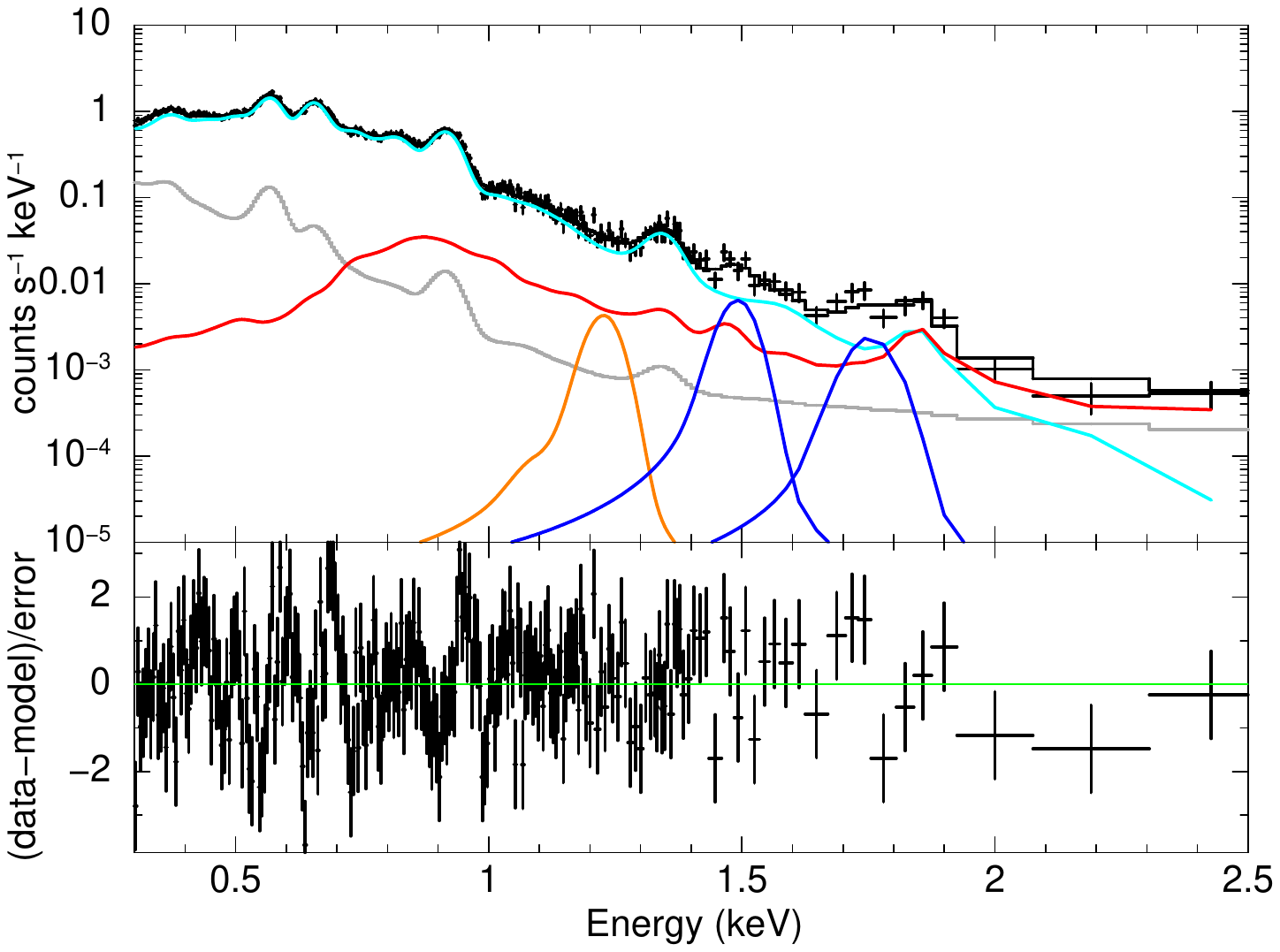}{0.18\textwidth}{layer 20}}
\gridline{\fig{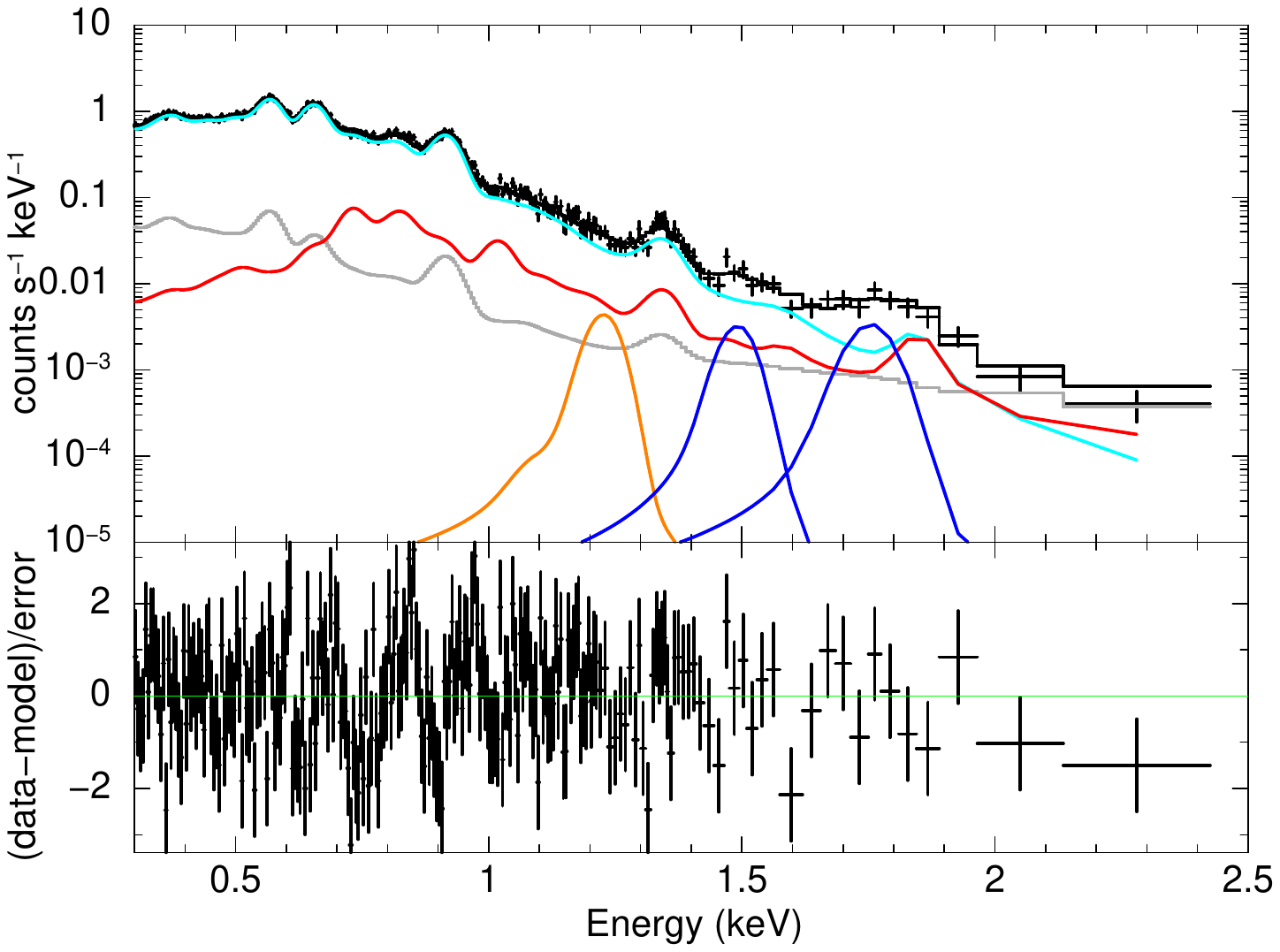}{0.18\textwidth}{layer 21}
          \fig{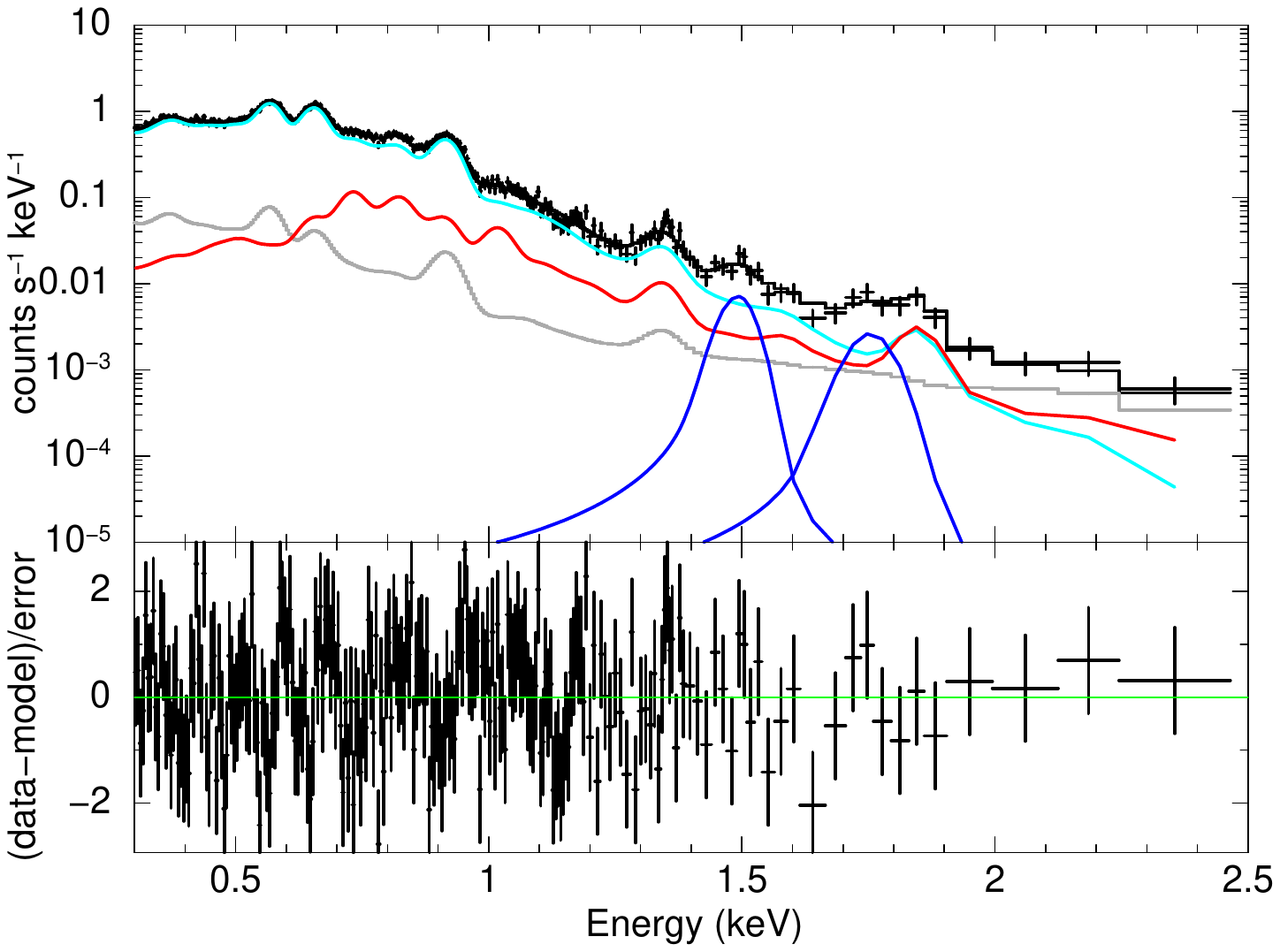}{0.18\textwidth}{layer 22}
          \fig{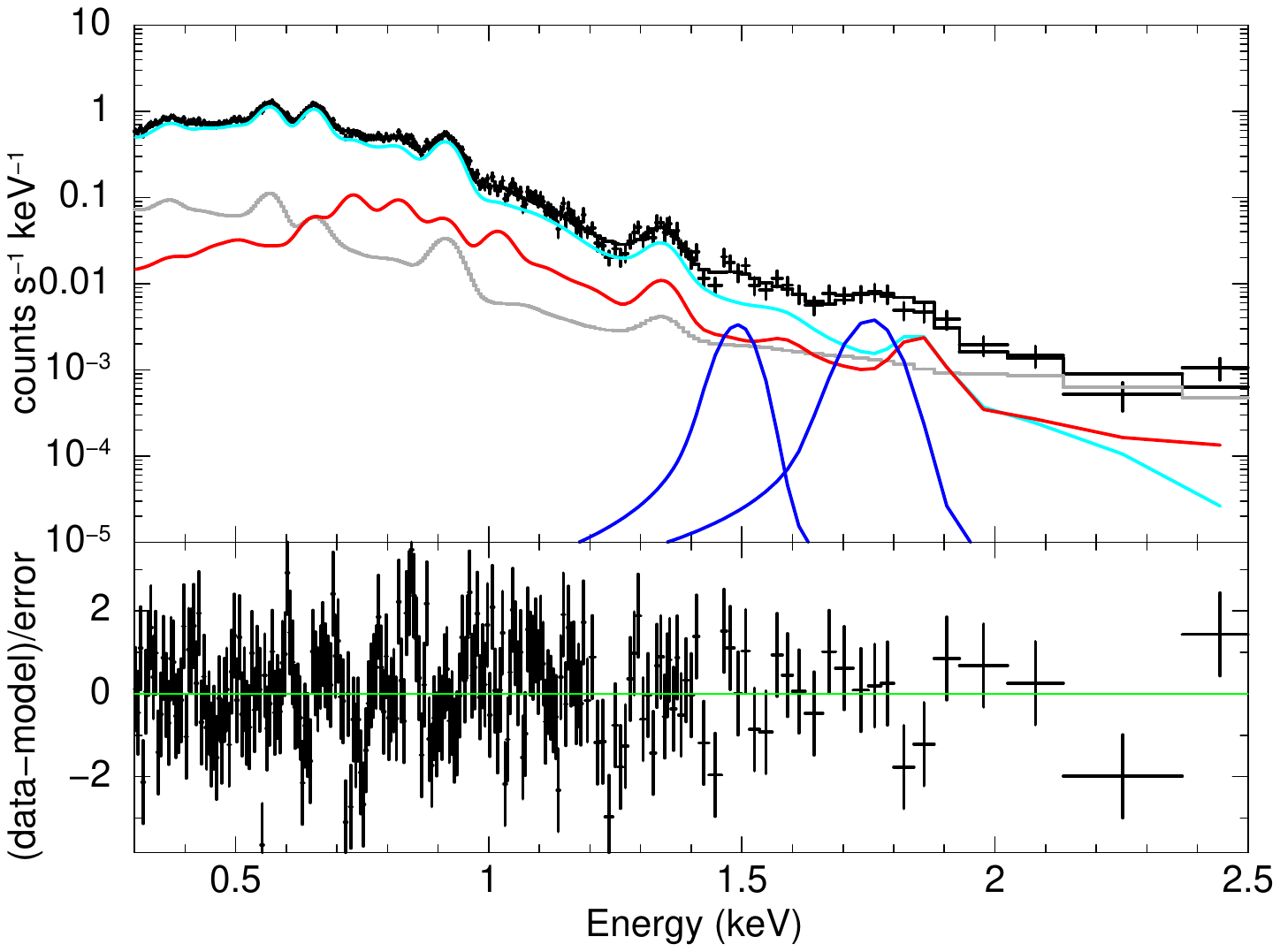}{0.18\textwidth}{layer 23}
          \fig{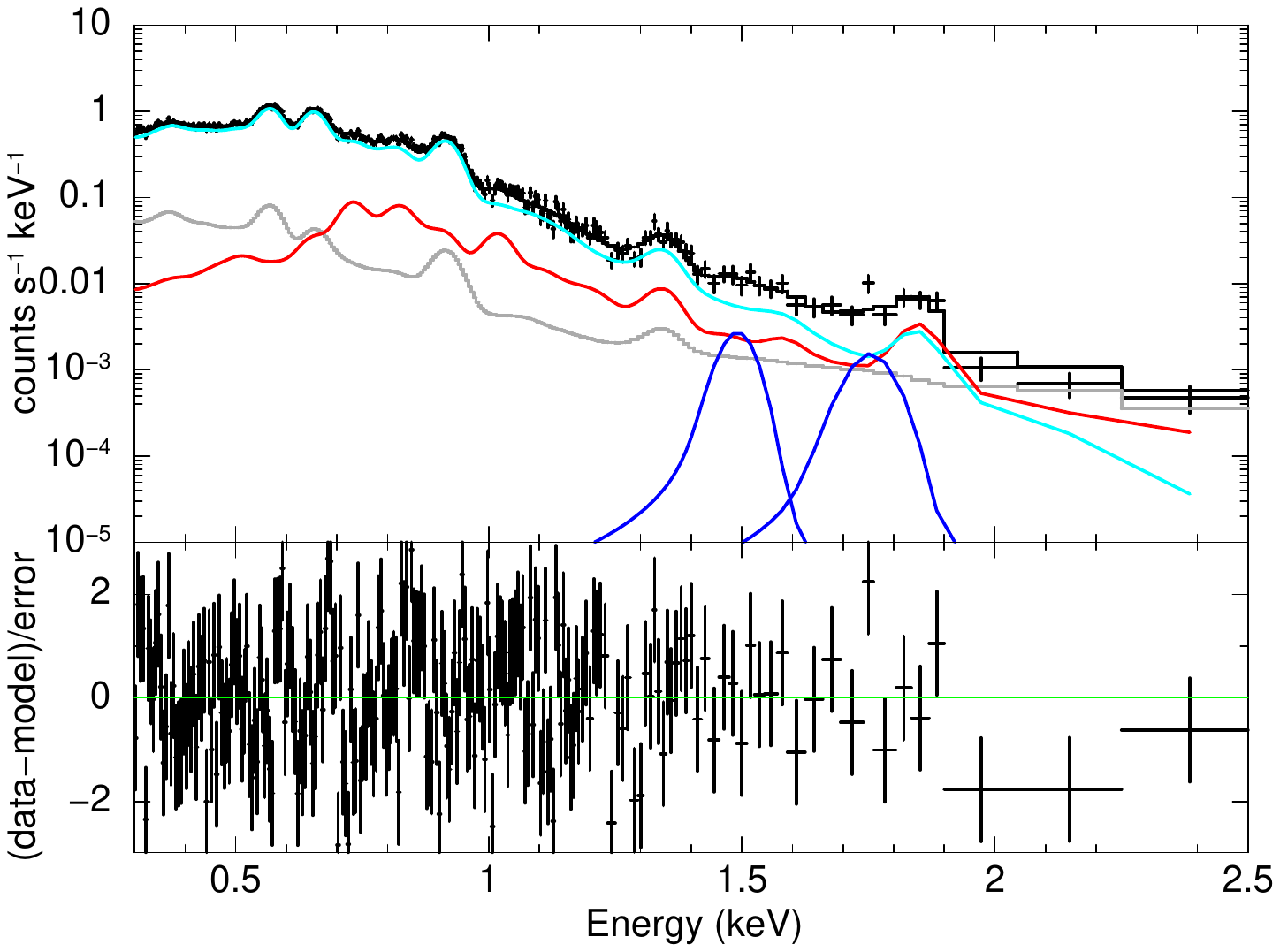}{0.18\textwidth}{layer 24}}
\caption{The spectra and best-fit model of Fan-shaped regions. Cyan, red, orange, gray, and blue lines show the model of the low-temperature component, the high-temperature component, Fe L line, the source background, and the NXB fluorescent lines, respectively. We adopt 1 {\tt nei} model to the spectra of layer 1--16 and 2 {\tt nei} model to the spectra of layer 17--24.\label{fig:best-fit_annulus}}
\end{figure}

\begin{figure}[ht!]
\plotone{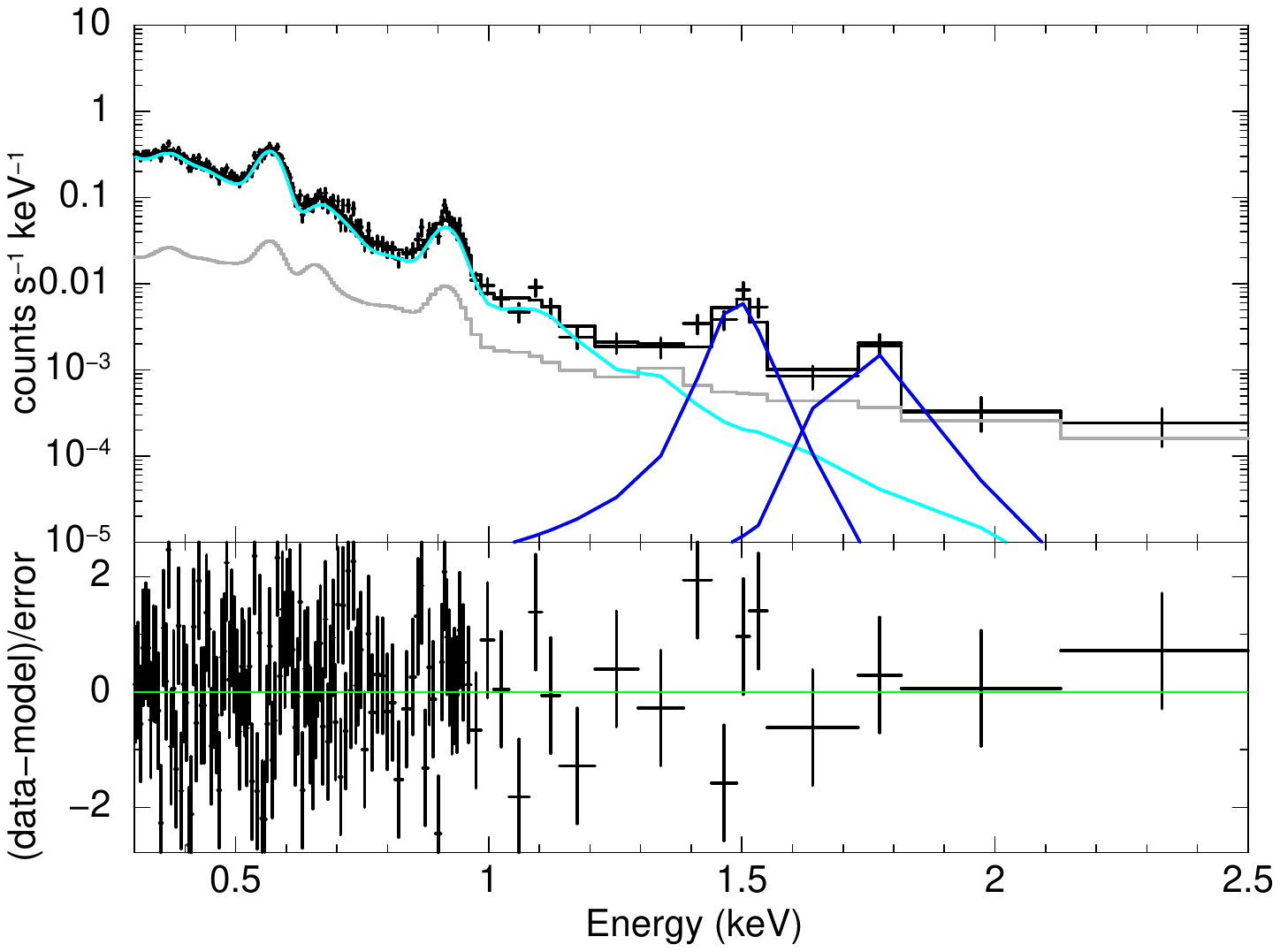}
\caption{The spectrum and best-fit model of outer shell region. Cyan, gray and blue line shows the source model, Fe L line, source background and NXB fluorescent lines, respectively. \label{fig:best-fit_outershell}}
\end{figure}

We fit the spectra of all 24 regions independently. The spectra of 16 outer regions (layer 1--16) and the outermost box region are well-fitted by a one temperature {\tt nei} model. On the other hand, the spectra of the 8 inner regions (layer 17--24) have some local residuals around O and Ne lines and their $\chi^2$ values are high with one temperature {\tt nei} model. 
These residuals can be reduced by applying two {\tt nei} components with different temperatures\footnote{For example, the residual of layer 17 with 1 {\tt nei} model is $\chi^2=308.93/216\;\mathrm{d.o.f}$, and that with 2 {\tt nei} model is $\chi^2=282.12/214\;\mathrm{d.o.f}$. The possibility of F-test is $6.04\times10^{-5}$.}. We also apply the same model to layer 1--16, but the emission measure of the additional components is consistent with zero value. We note that the addition of the hot component in inner regions does not influence the parameter value of the low-temperature component. All abundances and  $n_{\rm e}t$ are linked between two components. Two {\tt nei} components can adequately reproduce the spectra of the inner regions. 
The spectra and the best-fit models are shown in 
Figure~\ref{fig:best-fit_annulus} (fan-shaped regions)
and
Figure~\ref{fig:best-fit_outershell} (the outer shell region). 
The best-fit parameters are summerized in 
Table~\ref{tab:src_parameter_1nei} (the outer region with one temperature model), 
Table~\ref{tab:src_parameter_2ne1} (the inner region with two temperature model), 
and Table~\ref{tab:outershell_parmeter} (the outer shell region).

\begin{splitdeluxetable*}{ccccccBccccccBcccccc}
\tabletypesize{\scriptsize}
\tablewidth{0pt} 
\tablecaption{Best-fit parameter of source region (1 temperature model) \label{tab:src_parameter_1nei}}
\tablehead{
&&\multicolumn{4}{c}{layer}&\multicolumn{6}{c}{layer}&\multicolumn{6}{c}{layer}\\
\cline{3-18}
Model&Component&1&2&3&4&5&6&7&8&9&10&11&12&13&14&15&16}
\startdata 
tbabs&$N_\mathrm{H}$($\times10^{20}\;\mathrm{cm^{-2}}$)&\multicolumn{4}{c}{$6.0$ (fix)}&\multicolumn{6}{c}{$6.0$ (fix)}&\multicolumn{6}{c}{$6.0$ (fix)}\\
vnei&kT (keV)&$0.16_{-0.02}^{+0.03}$&$0.15_{-0.01}^{+0.02}$&$0.16_{-0.01}^{+0.02}$&$0.18_{-0.02}^{+0.01}$&$0.17_{-0.01}^{+0.01}$&$0.18_{-0.02}^{+0.01}$&$0.17_{-0.01}^{+0.02}$&$0.17_{-0.02}^{+0.01}$&$0.17_{-0.01}^{+0.01}$&$0.17_{-0.01}^{+0.01}$&$0.18_{-0.01}^{+0.01}$&$0.18_{-0.01}^{+0.01}$&$0.18_{-0.01}^{+0.01}$&$0.18_{-0.01}^{+0.01}$&$0.18_{-0.01}^{+0.01}$&$0.19_{-0.01}^{+0.01}$\\
&C (solar)&$0.14_{-0.10}^{+0.22}$&$0.18_{-0.10}^{+0.09}$&$0.46_{-0.19}^{+0.28}$&$0.18_{-0.09}^{+0.12}$&$0.32_{-0.12}^{+0.18}$&$0.34_{-0.14}^{+0.19}$&$0.42_{-0.17}^{+0.24}$&$0.29_{-0.10}^{+0.18}$&$0.17_{-0.07}^{+0.08}$&$0.23_{-0.07}^{+0.09}$&$0.33_{-0.10}^{+0.13}$&$0.53_{-0.16}^{+0.15}$&$0.59_{-0.20}^{+0.21}$&$0.82_{-0.13}^{+0.28}$&$0.84_{-0.31}^{+0.37}$&$0.99_{-0.35}^{+0.44}$\\
&N (solar)&$<0.06$&$0.10_{-0.04}^{+0.05}$&$0.13_{-0.04}^{+0.05}$&$0.14_{-0.04}^{+0.05}$&$0.15_{-0.04}^{+0.04}$&$0.18_{-0.04}^{+0.05}$&$0.23_{-0.05}^{+0.06}$&$0.13_{-0.03}^{+0.03}$&$0.12_{-0.02}^{+0.02}$&$0.16_{-0.02}^{+0.03}$&$0.15_{-0.02}^{+0.03}$&$0.22_{-0.04}^{+0.03}$&$0.26_{-0.04}^{+0.04}$&$0.30_{-0.05}^{+0.05}$&$0.35_{-0.07}^{+0.10}$&$0.36_{-0.07}^{+0.12}$\\
&O (solar)&$0.11_{-0.03}^{+0.04}$&$0.14_{-0.03}^{+0.03}$&$0.18_{-0.03}^{+0.04}$&$0.15_{-0.02}^{+0.03}$&$0.17_{-0.02}^{+0.03}$&$0.18_{-0.03}^{+0.03}$&$0.21_{-0.03}^{+0.04}$&$0.16_{-0.02}^{+0.02}$&$0.14_{-0.01}^{+0.02}$&$0.13_{-0.01}^{+0.01}$&$0.15_{-0.02}^{+0.02}$&$0.18_{-0.03}^{+0.02}$&$0.20_{-0.02}^{+0.02}$&$0.24_{-0.03}^{+0.03}$&$0.27_{-0.03}^{+0.04}$&$0.28_{-0.04}^{+0.04}$\\
&Ne (solar)&$0.18_{-0.07}^{+0.09}$&$0.32_{-0.07}^{+0.10}$&$0.36_{-0.07}^{+0.09}$&$0.32_{-0.06}^{+0.05}$&$0.41_{-0.06}^{+0.07}$&$0.38_{-0.07}^{+0.08}$&$0.53_{-0.10}^{+0.11}$&$0.33_{-0.05}^{+0.07}$&$0.29_{-0.03}^{+0.04}$&$0.31_{-0.03}^{+0.04}$&$0.32_{-0.03}^{+0.04}$&$0.39_{-0.05}^{+0.04}$&$0.45_{-0.05}^{+0.05}$&$0.53_{-0.06}^{+0.08}$&$0.59_{-0.08}^{+0.09}$&$0.61_{-0.07}^{+0.11}$\\
&Mg (solar)&$<0.10$&$<0.39$&$0.30_{-0.16}^{+0.20}$&$0.18_{-0.11}^{+0.16}$&$0.26_{-0.11}^{+0.11}$&$0.28_{-0.12}^{+0.15}$&$0.30_{-0.14}^{+0.16}$&$0.13_{-0.08}^{+0.10}$&$0.09_{-0.05}^{+0.05}$&$0.14_{-0.06}^{+0.06}$&$0.20_{-0.06}^{+0.07}$&$0.27_{-0.04}^{+0.07}$&$0.23_{-0.06}^{+0.07}$&$0.26_{-0.07}^{+0.08}$&$0.37_{-0.09}^{+0.12}$&$0.35_{-0.08}^{+0.08}$\\
&Fe (solar)&$0.18_{-0.16}^{+0.57}$&$0.44_{-0.30}^{+0.69}$&$0.29_{-0.18}^{+0.38}$&$0.20_{-0.08}^{+0.27}$&$0.24_{-0.11}^{+0.21}$&$0.22_{-0.08}^{+0.10}$&$0.54_{-0.28}^{+0.47}$&$0.16_{-0.05}^{+0.13}$&$0.20_{-0.06}^{+0.08}$&$0.15_{-0.03}^{+0.05}$&$0.20_{-0.05}^{+0.06}$&$0.26_{-0.07}^{+0.03}$&$0.30_{-0.08}^{+0.08}$&$0.34_{-0.08}^{+0.10}$&$0.34_{-0.09}^{+0.14}$&$0.35_{-0.08}^{+0.14}$\\
&nt ($\mathrm{s\;cm^{-3}}$)&$1.8_{-1.4}^{+3.5}\times10^{11}$&$2.5_{-1.3}^{+1.7}\times10^{11}$&$3.3_{-1.5}^{+1.7}\times10^{11}$&$1.8_{-0.5}^{+1.8}\times10^{11}$&$3.1_{-1.2}^{+1.5}\times10^{11}$&$2.5_{-0.7}^{+0.7}\times10^{11}$&$3.2_{-1.2}^{+2.0}\times10^{11}$&$2.5_{-0.6}^{+0.9}\times10^{11}$&$2.5_{-0.4}^{+0.7}\times10^{11}$&$2.6_{-0.4}^{+0.6}\times10^{11}$&$3.2_{-0.5}^{+0.6}\times10^{11}$&$3.8_{-0.8}^{+0.7}\times10^{11}$&$4.6_{-1.0}^{+1.0}\times10^{11}$&$5.3_{-1.2}^{+1.3}\times10^{11}$&$7.2_{-2.0}^{+5.7}\times10^{11}$&$6.3_{-1.6}^{+4.6}\times10^{11}$\\
&norm ($\mathrm{cm}^{-2}$)&$4.5_{-1.4}^{+1.2}\times10^{-3}$&$7.0_{-1.4}^{+1.5}\times10^{-3}$&$7.0_{-1.4}^{+1.4}\times10^{-3}$&$7.2_{-1.0}^{+1.6}\times10^{-3}$&$1.1_{-0.2}^{+0.2}\times10^{-2}$&$8.0_{-1.1}^{+1.1}\times10^{-3}$&$9.45_{-1.6}^{+1.6}\times10^{-3}$&$1.4_{-0.2}^{+0.2}\times10^{-2}$&$2.3_{-0.2}^{+0.3}\times10^{-2}$&$2.4_{-0.2}^{+0.2}\times10^{-2}$&$2.3_{-0.2}^{+0.2}\times10^{-2}$&$2.2_{-0.2}^{+0.2}\times10^{-2}$&$1.9_{-0.2}^{+0.2}\times10^{-2}$&$1.5_{-0.2}^{+0.2}\times10^{-2}$&$1.4_{-0.2}^{+0.2}\times10^{-2}$&$1.2_{-0.2}^{+0.2}\times10^{-2}$\\
gauss&norm (photon/keV/cm$^2$/s)&$<6.8\times10^{-7}$&$<1.8\times10^{-7}$&$<4.1\times10^{-7}$&$5.0_{-4.8}^{+7.0}\times10^{-7}$&$<1.4\times10^{-7}$&$<4.9\times10^{-7}$&$<1.0\times10^{-6}$&$<1.0\times10^{-6}$&$<8.1\times10^{-7}$&$<4.6\times10^{-7}$&$<1.1\times10^{-6}$&$<9.1\times10^{-7}$&$<3.3\times10^{-7}$&$<1.4\times10^{-6}$&$<7.0\times10^{-7}$&$<1.2\times10^{-6}$\\
\hline
$\chi^2$/d.o.f&&194.36/128&156.33/145&178.79/154&164.24/160&186.26/172&185.51/173&226.47/177&225.84/183&288.81/183&209.16/195&243.19/200&245.15/209&279.59/210&239.07/214&268.13/210&322.85/211\\
\enddata
\end{splitdeluxetable*}

\begin{splitdeluxetable*}{ccccccBccccccBcccccc}
\tabletypesize{\scriptsize}
\tablewidth{0pt} 
\tablecaption{Best-fit parameter of source region (2 temperature model)\label{tab:src_parameter_2ne1}}
\tablehead{
&&\multicolumn{4}{c}{layer}&\multicolumn{6}{c}{layer}&\multicolumn{6}{c}{layer}\\
\cline{3-18}
Model&Component&\multicolumn{2}{c}{17}&\multicolumn{2}{c}{18}&\multicolumn{2}{c}{19}&\multicolumn{2}{c}{20}&\multicolumn{2}{c}{21}&\multicolumn{2}{c}{22}&\multicolumn{2}{c}{23}&\multicolumn{2}{c}{24}\\
\cline{3-18}
&&cold&hot&cold&hot&cold&hot&cold&hot&cold&hot&cold&hot&cold&hot&cold&hot}
\startdata 
\hline
tbabs&$N_\mathrm{H}$($\times10^{20}\;\mathrm{cm^{-2}}$)&\multicolumn{4}{c}{$6.0$ (fix)}&\multicolumn{6}{c}{$6.0$ (fix)}&\multicolumn{6}{c}{$6.0$ (fix)}\\
vnei&kT (keV)&$0.18_{-0.02}^{+0.01}$&$0.29_{-0.13}^{+0.14}$&$0.18_{-0.01}^{+0.01}$&$0.48_{-0.13}^{+0.17}$&$0.19_{-0.01}^{+0.01}$&$0.34_{-0.07}^{+0.12}$&$0.19_{-0.01}^{+0.02}$&$0.78_{-0.20}^{+0.10}$&$0.19_{-0.01}^{+0.01}$&$0.43_{-0.08}^{+0.15}$&$0.19_{-0.01}^{+0.01}$&$0.36_{-0.06}^{+0.09}$&$0.20_{-0.01}^{+0.01}$&$0.35_{-0.03}^{+0.12}$&$0.19_{-0.01}^{+0.01}$&$0.41_{-0.10}^{+0.07}$\\
&C (solar)&\multicolumn{2}{c}{$0.90_{-0.28}^{+0.35}$}&\multicolumn{2}{c}{$1.7_{-0.7}^{+0.5}$}&\multicolumn{2}{c}{$1.0_{-0.5}^{+0.6}$}&\multicolumn{2}{c}{$2.0_{-0.7}^{+1.2}$}&\multicolumn{2}{c}{$2.0_{-0.6}^{+1.0}$}&\multicolumn{2}{c}{$1.3_{-0.6}^{+0.8}$}&\multicolumn{2}{c}{$2.0_{-0.8}^{+1.3}$}&\multicolumn{2}{c}{$0.92_{-0.54}^{+0.70}$}\\
&N (solar)&\multicolumn{2}{c}{$0.53_{-0.08}^{+0.05}$}&\multicolumn{2}{c}{$0.62_{-0.06}^{+0.09}$}&\multicolumn{2}{c}{$0.55_{-0.10}^{+0.11}$}&\multicolumn{2}{c}{$0.80_{-0.16}^{+0.17}$}&\multicolumn{2}{c}{$0.78_{-0.07}^{+0.17}$}&\multicolumn{2}{c}{$0.51_{-0.10}^{+0.09}$}&\multicolumn{2}{c}{$0.73_{-0.08}^{+0.23}$}&\multicolumn{2}{c}{$0.48_{-0.07}^{+0.14}$}\\
&O (solar)&\multicolumn{2}{c}{$0.36_{-0.03}^{+0.06}$}&\multicolumn{2}{c}{$0.44_{-0.07}^{+0.06}$}&\multicolumn{2}{c}{$0.39_{-0.04}^{+0.06}$}&\multicolumn{2}{c}{$0.47_{-0.04}^{+0.09}$}&\multicolumn{2}{c}{$0.47_{-0.07}^{+0.07}$}&\multicolumn{2}{c}{$0.41_{-0.05}^{+0.08}$}&\multicolumn{2}{c}{$0.49_{-0.07}^{+0.12}$}&\multicolumn{2}{c}{$0.38_{-0.05}^{+0.07}$}\\
&Ne (solar)&\multicolumn{2}{c}{$0.73_{-0.08}^{+0.10}$}&\multicolumn{2}{c}{$0.90_{-0.15}^{+0.11}$}&\multicolumn{2}{c}{$0.72_{-0.09}^{+0.13}$}&\multicolumn{2}{c}{$0.91_{-0.13}^{+0.13}$}&\multicolumn{2}{c}{$0.87_{-0.13}^{+0.13}$}&\multicolumn{2}{c}{$0.72_{-0.09}^{+0.14}$}&\multicolumn{2}{c}{$0.82_{-0.12}^{+0.24}$}&\multicolumn{2}{c}{$0.72_{-0.10}^{+0.12}$}\\
&Mg (solar)&\multicolumn{2}{c}{$0.49_{-0.10}^{+0.13}$}&\multicolumn{2}{c}{$0.61_{-0.17}^{+0.12}$}&\multicolumn{2}{c}{$0.47_{-0.10}^{+0.13}$}&\multicolumn{2}{c}{$0.75_{-0.13}^{+0.16}$}&\multicolumn{2}{c}{$0.66_{-0.11}^{+0.11}$}&\multicolumn{2}{c}{$0.44_{-0.10}^{+0.14}$}&\multicolumn{2}{c}{$0.63_{-0.14}^{+0.22}$}&\multicolumn{2}{c}{$0.42_{-0.11}^{+0.12}$}\\
&Fe (solar)&\multicolumn{2}{c}{$0.57_{-0.13}^{+0.12}$}&\multicolumn{2}{c}{$0.68_{-0.13}^{+0.10}$}&\multicolumn{2}{c}{$0.51_{-0.07}^{+0.11}$}&\multicolumn{2}{c}{$0.71_{-0.12}^{+0.14}$}&\multicolumn{2}{c}{$0.66_{-0.11}^{+0.13}$}&\multicolumn{2}{c}{$0.53_{-0.07}^{+0.12}$}&\multicolumn{2}{c}{$0.59_{-0.09}^{+0.20}$}&\multicolumn{2}{c}{$0.51_{-0.09}^{+0.08}$}\\
&nt ($\mathrm{s\;cm^{-3}}$)&\multicolumn{2}{c}{$>7.4\times10^{11}$}&\multicolumn{2}{c}{$>1.1\times10^{12}$}&\multicolumn{2}{c}{$>1.3\times10^{12}$}&\multicolumn{2}{c}{$>7.1\times10^{11}$}&\multicolumn{2}{c}{$>1.9\times10^{12}$}&\multicolumn{2}{c}{$>1.4\times10^{12}$}&\multicolumn{2}{c}{$>1.6\times10^{12}$}&\multicolumn{2}{c}{$>1.6\times10^{12}$}\\
&norm 
($\mathrm{cm}^{-2}$)&$1.0_{-0.3}^{+0.2}\times10^{-2}$&$0.4_{-0.3}^{+2.4}\times10^{-3}$&$8.0_{-0.9}^{+1.5}\times10^{-3}$&$7.2_{-3.2}^{+8.0}\times10^{-5}$&$8.4_{-0.6}^{+1.1}\times10^{-3}$&$2.6_{-1.7}^{+3.5}\times10^{-4}$&$6.5_{-1.1}^{+1.1}\times10^{-3}$&$4.4_{-1.5}^{+2.4}\times10^{-5}$&$6.8_{-1.0}^{+1.1}\times10^{-3}$&$1.2_{-0.6}^{+4.3}\times10^{-4}$&$7.1_{-1.5}^{+0.8}\times10^{-3}$&$2.8_{-1.6}^{+3.1}\times10^{-4}$&$5.4_{-1.0}^{+0.7}\times10^{-3}$&$2.4_{-1.7}^{+4.4}\times10^{-4}$&$6.7_{-1.0}^{+1.0}\times10^{-3}$&$1.9_{-1.0}^{+2.9}\times10^{-4}$\\
gauss&norm (photon/keV/cm$^2$/s)&\multicolumn{2}{c}{$<1.1\times10^{-6}$}&\multicolumn{2}{c}{$6.1<\times10^{-7}$}&\multicolumn{2}{c}{$<6.8\times10^{-7}$}&\multicolumn{2}{c}{$<2.2\times10^{-6}$}&\multicolumn{2}{c}{$<2.2\times10^{-6}$}&\multicolumn{2}{c}{$<7.6\times10^{-7}$}&\multicolumn{2}{c}{$<2.4\times10^{-7}$}&\multicolumn{2}{c}{$<7.3\times10^{-7}$}\\
\hline
$\chi^2$/d.o.f&&\multicolumn{2}{c}{282.12/214}&\multicolumn{2}{c}{270.01/218}&\multicolumn{2}{c}{252.15/223}&\multicolumn{2}{c}{300.41/222}&\multicolumn{2}{c}{301.39/219}&\multicolumn{2}{c}{256.29/220}&\multicolumn{2}{c}{321.93/220}&\multicolumn{2}{c}{306.21/213}\\
\enddata
\end{splitdeluxetable*}

\begin{table}[ht]
    \centering
    \caption{Best-fit parameter for the outer shell region}
    \label{tab:outershell_parmeter}
    \begin{tabular}{ccc}
    \hline
         \multicolumn{2}{c}{Parameter}&Value  \\
         \hline
         tbabs & $N_\mathrm{H}$($\times10^{20}\;\mathrm{cm^{-2}}$)&$6.0$(fix)\\
         \multirow{4}{*}{vnei}&kT(keV)&$0.15_{-0.01}^{+0.03}$\\
         &C&$0.11_{-0.05}^{+0.13}$\\
         &N&$0.03_{-0.03}^{+0.04}$\\
         &O&$0.11_{-0.02}^{+0.03}$\\
         &Ne&$0.23_{-0.07}^{+0.12}$\\
         &Mg&$<0.29$\\
         &Fe&$0.48_{-0.36}^{+1.28}$\\
         &nt ($\mathrm{s\;cm^{-3}}$)&$1.3_{-1.2}^{+2.8}\times10^{11}$\\
         &norm ($\mathrm{cm}^{-2}$)&$5.5_{-2.3}^{+1.7}\times10^{-3}$\\
         gauss&norm (photon/keV/cm$^2$/s)&$<2.9\times10^{-7}$\\
         \hline
         \multicolumn{2}{c}{$\chi^2$/d.o.f}&168.48/125\\
         \hline
    \end{tabular}
\end{table}

\section{Discussion}
In this section, we discuss the value and distribution of both the electron temperature and $n_{\rm e}t$. 
Our observational results are compared with previous studies and the theoretical values
in section 4.1 and 4.2, respectively.
Section 4.3 and 4.4 address the variation of the electron temperature and $n_{\rm e}t$, respectively. 

\subsection{The comparison to previous studies}
\begin{figure}[ht!]
\plottwo{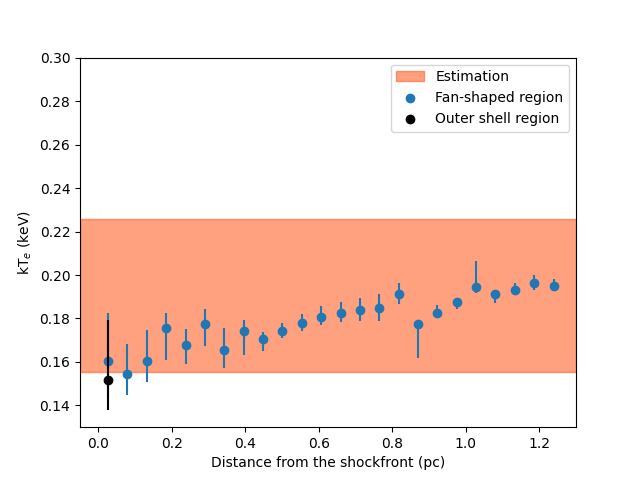}{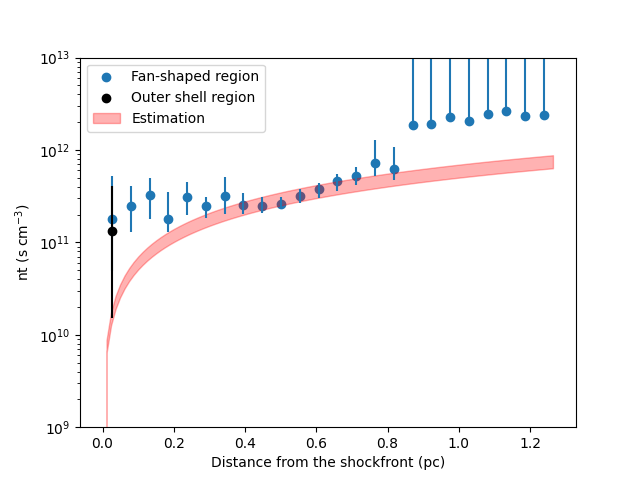}
\caption{The spatial distribution of kT (left) and $n_{\rm e}t$ (right) of the low-temperature component in the northeastern region of the Cygnus Loop. The blue points and the black point show the result of the annulus regions and the outer shell region, respectively.
The red region in the left figure shows the range of an electron temperature estimated from both a proper motion and the distance of the Cygnus Loop. The red region of the right figure shows the range of a $n_{\rm e}t$ estimation from an electron density and a shock velocity of the Cygnus Loop.\label{fig:nt-kT_distribution}} 
\end{figure}

Figure~\ref{fig:nt-kT_distribution} shows the distribution of $T_e$ (left panel) and $n_{\rm e}t$ (right panel) of the low-temperature component. 
The electron temperature shows an increasing trend inward from 0.15 to 0.19~keV over $6\;\mathrm{arcmin}$ (or $1.27\;\mathrm{pc}$ at a distance $d=725$~pc). 
The value of $T_e$ in the low-temperature component suddenly decreases at 0.87~pc from the shock. 
Because we introduce another high-temperature component inside this point, the systematic errors caused by model uncertainty may contribute to this decrease.
The temperature value (the low-temperature component in 2 {\tt nei} model) is higher than that reported by \cite{2007PASJ...59S.163M}, 
and lower than that of the high-temperature component in \cite{2007PASJ...59S.163M} and \cite{2008ApJ...680.1198K}. 
The high-temperature component in layer 17--24 shows similar behavior to the result in  \cite{2007PASJ...59S.163M} except for layer 20. 
The value of $n_{\rm e}t$ is constant in the outer region, which is coincident  with the H$_\alpha$ shell region. 
It increases inward in the region inside the H$_\alpha$ shell (the region with a distance of more than 0.4 pc away from the shock front). 
The value of $n_{\rm e}t$ in \cite{2007PASJ...59S.163M} is also similar to ours, however, the temperature distribution derived by \cite{2007PASJ...59S.163M} is roughly flat, which is not consistent with our result. 

\subsection{The comparison between the theoretical and observed values of the electron temperature and \texorpdfstring{$n_{\rm e}t$}{ne t}}

The ion temperature and $n_{\rm e}t$ is theoretically calculated by simple hydrodynamics with environmental parameters. 
In this section, we compare the theoretical values with our analysis results.

The ion temperature just behind the shock is given by:
\begin{equation}
kT_2=mv_1^2\cdot \frac{(\gamma-1)M_1^2+2}{(\gamma+1)M_1^2}\left(\frac{1}{\gamma M_1}+\frac{2}{\gamma+1}-\frac{2}{M_1^2(\gamma+1)}\right),
\end{equation}
%
where subscripts 1 and 2 denote the quantities upstream and downstream of the shock, respectively. The upstream velocity $v_1$ is equivalent to a shock velocity of the Cygnus Loop, which has been measured from the proper motion to be 3.4--4.1~arcsec in 39.1~yr \citep{2009ApJ...702..327S}. 
For the distance 
$d=725\pm15\;\mathrm{pc}$ \citep{2021MNRAS.507..244F}, 
the shock velocity is 310--380~km~s$^{-1}$. 
Quantities $m$ and $\gamma$ denote the particle mass and an adiabatic index, respectively. The sonic Mach number is calculated as $M_1\equiv v_1\sqrt{\rho_1/\gamma p_1}$, where $\rho_1$ and $p_1$ denote upstream gas pressure and  gas density, respectively. 
The sound speed in the Galactic plane is typically $\sim20\;\mathrm{km\;s^{-1}}$ \citep{2024A&A...687A.127M}. 
Using Equation~(1), we can estimate the proton temperature just behind the shock to be 0.19--0.28~keV.
For slowly moving shock front such as the Cygnus Loop, the electron temperature increases due to the collisionless heating, 
so that the relation between the electron and the proton temperatures is not mass-proportional. Some previous studies estimated this temperature ratio: $kT_{\rm e}/kT_{\rm p}=0.67-1.0$ in \cite{2001ApJ...547..995G} and  $kT_{\rm e}/kT_{\rm p}=0.8-1.0$ in \cite{2023ApJ...949...50R}. 

The value of $n_{\rm e}t$ can be estimated as the product of the electron density and the time after the shock passage. 
The pre-shock electron density has been estimated as $0.4\pm0.1\;\mathrm{cm^{-3}}$ \citep{2003ApJ...584..770R}, 
leading to the post-shock density of $1.6\pm0.4\;\mathrm{cm^{-3}}$.
The time after the shock passage is estimated by dividing the distance from the shock by a quarter of the shock velocity. 
For example, 
since the center of the outer shell region in Figure~\ref{fig:region_all} 
is located at 15~arcsec from the shock front, the elapsed time is calculated as (0.2--$2.1)\times10^{10}$~s, so that the value of $n_{\rm e}t$ at this position is estimated to be (2.4--$4.2)\times10^{10}$~s~cm$^{-3}$.

We compare these estimations and our analysis results in 
Figure~\ref{fig:nt-kT_distribution}. 
The temperature ratio in the outer shell region is $kT_{\rm e}/kT_{\rm p}=0.54$--0.80.
This is consistent with the value in previous studies 
\citep{2001ApJ...547..995G,2023ApJ...949...50R}.
As for the $n_{\rm e}t$, it is greater than the estimation of the electron density and the time just behind the shock, even we assume the upper end of the post-shock electron density, $n_{\rm e}\approx2.0\;\mathrm{cm^{-3}}$, inferred by \citet{2003ApJ...584..770R}. 
These values differ by about an order of magnitude, which requires some processes that rapidly proceed the ionization states of the plasma.

\subsection{The origin of the electron temperature variation}
In our analysis, the electron temperature far downstream is significantly higher than that just behind the shock. A similar trend of the electron temperature can be shown in previous studies such as \cite{2007PASJ...59S.163M}, but the temperature variation scale in our result is smaller than that in previous studies.

There are some possibilities to explain the increase in electron temperature. First, we consider the electron temperature variation in our result by Coulomb relaxation. However, the electron temperature distribution in Coulomb relaxation is almost constant because the electron and the proton temperature is close just behind the shock.

\begin{figure}[ht!]
\plotone{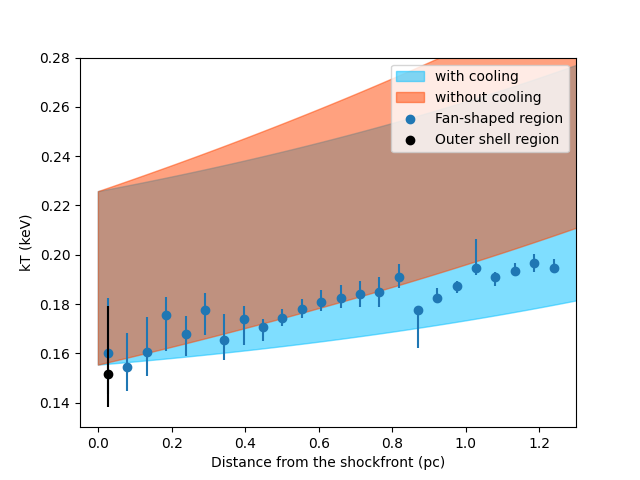}
\caption{The comparison of the temperature in our result to the temperature structure of the Sedov solution. The red region shows the case without radiative cooling, and the blue region shows the case with radiative cooling of the cooling function in \cite{2023ApJS..264...10K}. The brown region is the overlapping region of both cases.
\label{fig:kT_model_comparison}}
\end{figure}

We also consider the possibility of reflecting the internal temperature structure of the Cygnus Loop, which is described by the Sedov solution. In this case, we consider that our X-ray spectra are dominated by the emission from the heated circumstellar materials. Assuming the adiabatic expansion, 
we can estimate the distribution of pressure and density of plasma inside the shock as the Sedov solution. 
The temperature of a single fluid can be written as $kT\approx p/ n$. 
We assume $kT_{\rm e}/kT_{\rm p}=0.8$ just behind the shock front, taking into account previous studies \citep{2001ApJ...547..995G,2023ApJ...949...50R}.
The red region in Figure~\ref{fig:kT_model_comparison} shows the range of the electron temperature distribution 
inferred by the Sedov solution. Here, the radius of the Cygnus Loop, $R$, is assumed to be $19\;\mathrm{pc}$ \citep{1997ApJ...484..304L}. The spatial profile of the single fluid temperature is consistent with our result, and the slope is similar, although the data is on the lower end of the Sedov model (red region).

In the discussion above, we considered the effect derived from both the decrease of the shock velocity following the Sedov solution and the adiabatic expansion. However, there should be some energy loss effect. Some previous studies, such as \cite{2023ApJS..264...10K} and \cite{2016ApJ...821...20K}, calculated a cooling function $\Lambda_c$ under different conditions. \cite{2023ApJS..264...10K} calculated $\Lambda_c$ of ISM in collisional ionization equilibrium conditions, so the effects of dust heating and destruction due to shock passage were not considered. \cite{2016ApJ...821...20K} considered infrared dust emission due to the shock heating. However, they assumed that most of the heavy elements were contained in dust grains, leading to a low value of $\Lambda_c$ in the X-ray band. 
For this reason, the total value of $\Lambda_c$ from \cite{2016ApJ...821...20K} is lower than that from \cite{2023ApJS..264...10K}.
In order to consider the case with most significant cooling, 
we use $\Lambda_c$ in \cite{2023ApJS..264...10K} in the following. 
%
%
Then, using our analysis result of $\sim0.18$~keV, we get 
$\Lambda_c=6.2\times10^{-24}\;\mathrm{keV\;cm^{3}\;s^{-1}}$. 
Assuming a density of $n_{\rm e}=1.6\;\mathrm{cm^{-3}}$ \citep{2003ApJ...584..770R}, we estimate the cooling time of  $1.1\times10^5\;\mathrm{yr}$, about 10 times larger than the age of the Cygnus Loop, $\sim2\times10^4\;\mathrm{yr}$ \citep{2021MNRAS.507..244F}. This result indicates that the energy in our analysis region can be dissipated by about 10 \%. 
If the density is lower than that we assumed above, the value of $\Lambda_c$ becomes lower and the cooling time is longer.
The blue region of Figure~\ref{fig:kT_model_comparison} shows the temperature structure of the Sedov solution with radiative cooling of $\Lambda_c$ discussed above. This region is consistent with the temperature distribution in our analysis result.  
Note that the Sedov solution is derived on the assumption of the adiabatic expansion. 
It is difficult to strictly be applied in the region where radiative cooling is not ignored, that has been discussed here. 
However, in our model, it is just $\sim10\;\%$ of the total energy has been lost by cooling, which is still small compared with the total energy.

Another possible factor of the cooling is the collision with a high-density region such as the HI region. 
Our analysis region has some overlapping H$\alpha$ filaments. This overlapping may be caused by the collision with the high-density region. 
If this is true, the shock velocity will decrease, 
and the electron temperature will become lower than before. For this reason, the Sedov solution may no longer be applicable just behind the shock. 
Furthermore, the thermal conduction between the hot shock-heated plasma and the cold high-dense ISM will occur. This energy leakage lowers the post-shock temperature. 
Some previous studies mentioned the thermal conduction between the downstream shocked material and the upstream neutral gas or the molecular clouds as the origin of the recombination plasma 
\citep[e.g.,][]{2002ApJ...572..897K,2005ApJ...631..935K,Okon_2020,Sano_2021}.

We also consider the thermal conduction from the hot plasma in the inner regions. As shown in Figure~\ref{fig:kT_model_comparison} and discussed above, the electron temperature increases toward the interior. This temperature gradient causes the thermal conduction between the inner hot plasma and the outer cold gas, leading to an increase in the temperature just behind the shock. To evaluate the influence of thermal conduction, we compare the timescale of the thermal conduction to the dynamical timescale in our analysis region. The dynamical timescale $t_d$ is estimated from the length of our analysis region, $\sim1.2\;\mathrm{pc}$, and the shock velocity, $\sim300\;\mathrm{km\;s^{-1}}$, as $t_d\sim1.2\times10^{11}\;\mathrm{s}$.
The timescale of the thermal conduction $t_c$ is written as:
\begin{equation}
t_c\sim \frac{nkT(\delta x)^2}{\kappa \delta T},
\end{equation}
where $n$, $T$, $\kappa$, $\delta x$, and
$\delta T$ are the electron density, the electron temperature, the
thermal conductivity, the length scale, and the scale of the
temperature variation, respectively. 
\cite{1962pfig.book.....S} gave the form of  $\kappa$ as:

\begin{equation}
\kappa=\frac{1.84\times10^{-5}T^{5/2}}{\ln\Lambda}\;\mathrm{erg\;s^{-1}\;K^{-1}\;cm^{-1}},
\end{equation}
where the Coulomb logarithm, $\ln \Lambda$,  is approximated as:
\begin{equation}
\ln \Lambda=29.7+\ln n^{-1/2}(T/10^6\;\mathrm{K}),
\end{equation}
for $T>4.2\times10^5\;\mathrm{K}$ \citep{1977ApJ...211..135C}.
\noindent Considering the value of parameters in our analysis region, we estimate the thermal conductivity and the timescale of the thermal conduction as $\kappa\approx3.9\times10^{9}\;\mathrm{erg\;s^{-1}\;K^{-1}\;cm^{-1}}$ and $t_c\approx3.5\times10^{12}\;\mathrm{s}$. 
The timescale of the thermal conduction is significantly larger than the dynamical timescale, so that the thermal conduction from inner regions is small. We note that the thermal conductivity we use above does not consider the magnetic field. The magnetic field reduces the thermal conductivity perpendicular to the magnetic field \citep{1965RvPP....1..205B}, leading less significant thermal conduction from inner regions.

\subsection{The origin of \texorpdfstring{$n_{\rm e}t$}{ne t} variation}
As shown in section 4.1 and 
the right panel of Figure~\ref{fig:nt-kT_distribution}, 
the value of $n_{\rm e}t$ just behind the shock in our result is significantly larger than the value estimated from the product of the electron density and the time after a shock has passed. A similar $n_{\rm e}t$ excess is also observed in \cite{2007PASJ...59S.163M} which analyzed the same region of the Cygnus Loop. 

The ratio between OVII and OVIII is mainly determined by the $n_{\rm e}t$ values, but other factors such as resonance line scattering and charge exchange X-rays also influence this ratio. The self-absorption of the resonance scattering of OVII is larger than that of OVIII \citep{2008PASJ...60..521M}, leading to the overestimation of OVIII/OVII and the $n_{\rm e}t$ values. However, we cannot estimate how this effect influences the estimation of the $n_{\rm e}t$ values quantitatively. Therefore, we use the best-fit values of $n_{\rm e}t$ in our analysis as a reference value in the discussion below.

We discuss two possibilities for causing the excess of the $n_{e}t$ values just behind the shock.
The first possibility is due to the accelerated particles around the shock. They can interact and promote the ionization of the surrounding thermal plasma, leading to the observation of such $n_{\rm e}t$ excess \citep{2009ApJ...696.1956P}. However, as shown in section 4.1, the shock velocity of the Cygnus Loop is $\sim300\;\mathrm{km\;s^{-1}}$. It is unclear whether the statistical acceleration with such a slow shock can promote the value of $n_{\rm e}t$ to that of our result within the spatial scale of our analysis region.

The second possibility is the transport of the highly ionized plasma from the inner region. The plasma in the inner region is more ionized than in the region near the shock wave because more time has passed since the shock wave passed. By transporting this ionized plasma close to the shock, such as by turbulence, the values of $n_{\rm e}t$ values near the shock can increase. This turbulence might also produce the observed rippling of the shock front. In addition, the turbulence changes the downstream temperature distribution. However, as shown in Figure~\ref{fig:kT_model_comparison}, the temperature difference within $0.15\;\mathrm{pc}$ of the shock front is very small for the Sedov model. 
Hence, the electron temperature does not change significantly, which is consistent with our analysis results. Future high energy-resolution observations would verify whether the turbulence discussed above exists in our analysis region.

Note that the above discussion is qualitative and not directly observational. The simplest and most reliable way to discuss the ionization states is to directly analyze the ionization states of each ion from the emission lines in the spectra. If the $n_{\rm e}t$ increase is due to turbulence, both highly ionized and low-ionized emission lines are considered to coexist over the entire analysis region. On the other hand, if the shocked plasma is highly ionized just behind the shock, only the highly ionized emission lines can be observed. We could not analyze the ionization state of emission lines from the data of XMM-Newton due to the lack of energy resolution. With the currently operating satellites, XRISM \citep{2025PASJ..tmp...28T} can observe the ionization states on a large scale, but cannot analyze the spatial distribution as in this study. We look forward to the launch of future satellites with both high energy and spatial resolution, such as NewAthena.

\section{Conclusion}
We analyzed the northeastern region of the the Cygnus Loop using XMM-Newton and measured the spatial variation in electron temperature and ionization parameter just behind the shock. We divide the region into 24 fan-shaped sectors with a thickness of 15 arcsec (or $0.05 \;\mathrm{pc}$) and set one of the boundaries along the innermost H$_\alpha$ filament. We also analyze the outer shell region between two outermost H$_\alpha$ filaments to analyze the environment just behind the shock.
The electron temperature increases inward from $0.15\;\mathrm{keV}$ to $0.19\;\mathrm{keV}$. This value and distribution can be explained by the Sedov solution with radiative cooling. Ionized timescale $n_{\rm e}t$ is an order of magnitude larger than the estimation from the electron density and shock velocity. This high value is derived from the mixing with inner ionized plasma due to turbulence. Part of our discussion is qualitative and not based on direct observation, so we expect future satellites with both high energy and spatial resolution, such as NewAthena, to observe and analyze the ionization environment immediately after the shock in more detail.

\section{Acknowledgment}
This work was supported by JST SPRING Grant No.~JPMJSP2108 (MI), Grant-in-Aid for JSPS Fellows Grant No.~25KJ0968 (MI), and partially supported by Japan Society for the Promotion of Science Grants-in-Aid for Scientific Research (KAKENHI) 
Grant No.~JP23H01211, JP23K25907, JP23H04891, JP23H04895 (AB), 
23K22522, 23H04899 (RY), 
JP21H04487, JP24H01805, JP25K00999 (YO). 
We thank Dr.~Milica Andjelic for sharing H$_\alpha$ image of the Cygnus Loop. This data was collected with 2-m RCC telescope at Rozhen National Astronomical Observatory. 
We are also grateful for a careful reading of the manuscript and critical comments for its improvement to our referee, John~Raymond.

\begin{acknowledgments}
\end{acknowledgments}

%

\vspace{5mm}








\bibliography{cygnus_MI}{}
\bibliographystyle{aasjournal}



\end{document}